\documentclass[pra,twocolumn,groupaddress,graphics,amssymb,amsmath,superscriptaddress]{revtex4}
\usepackage{graphics}

\begin{document}

\title{Coherent control of trapped ions using off-resonant lasers}

\author{J. J. \surname{Garc\'{\i}a-Ripoll}}
\email{Juan.Ripoll@mpq.mpg.de}
\affiliation{Max-Planck-Institut f\"ur Quantenoptik, Hans-Kopfermann-Str. 1,
  Garching, D-85748, Germany.}
\author{P. \surname{Zoller}}
\affiliation{Institute for Theoretical Physics, University of
Innsbruck, and Institute for Quantum Optics and Quantum
Information of the Austrian Academy of Sciences, A-6020 Innsbruck,
Austria.}
\author{J. I. \surname{Cirac}}
\affiliation{Max-Planck-Institut f\"ur Quantenoptik, Hans-Kopfermann-Str. 1,
  Garching, D-85748, Germany.}


\date{\today}

\begin{abstract}
  In this paper we develop a unified framework to study the coherent
  control of trapped ions subject to state-dependent forces. Taking
  different limits in our theory, we can reproduce two different designs
  of a two-qubit quantum gate ---the pushing gate \cite{cirac00} and the
  fast gates based on laser pulses from Ref. \cite{ripoll03}---, and propose
  a new design based on continuous laser beams. We demonstrate how to
  simulate Ising Hamiltonians in a many ions setup, and how to create
  highly entangled states and induce squeezing. Finally, in a detailed
  analysis we identify the physical limits of this technique and study
  the dependence of errors on the temperature.
\end{abstract}

\maketitle

\section{Introduction}

Trapped ions constitute one of the most promising systems to
implement a scalable quantum computer \cite{levi03}. In such a
computer, information is stored in long-lived atomic states, and a
universal set of gates is obtained by manipulating these states
with lasers, and entangling the ions via the vibrational modes
\cite{cirac00}. During the last years we have seen experimental
demonstrations of various two--qubit gates
\cite{monroe95,demarco02,leibfried03,schmidtkaler03}, and it
remains to implement a robust scheme for scalability. Current
visions of a scalable computer are based on the idea of moving the
ions out of their storage area (or quantum memory) and make them
interact in pairs, performing two-qubit gates
\cite{cirac00,kielpinski02}.  Basic steps towards the experimental
implementation of this idea have been already demonstrated.
\cite{rowe02}.

However, outside quantum computing there are other promising areas
in which trapped ions may be of use.  On the one hand, there is a
great interest in preparing highly entangled or squeezed states,
which can be used both for metrology \cite{wineland92,wineland94},
or --- in a more fundamental fashion --- to characterize their
decoherence properties. The entanglement of ions is covered in a
variety of theoretical papers
\cite{steinbach98,molmer99,sorensen00,unanyan03}, and it has been
experimentally demonstrated for small systems
\cite{turchette98,sackett00,meyer01}. On the other hand, it has
also been suggested that ion traps can be used to simulate a
various spin Hamiltonians, ranging from local to long-range
interactions \cite{porras04,barjaktarevic}.

In all of these tasks ---quantum computing, creation of entanglement and
quantum simulation---, the goal is to induce some unitary evolution on the
internal state of the ions, which is used to store the information. For
instance, in the case of quantum computing it suffices to realize a phase gate
between two ions
\begin{equation}
  \label{phase-gate}
  U_{ph}(T) = e^{i\pi \sigma^z_1 \sigma^z_2/4},
\end{equation}
while in the case of quantum simulation we want a more general evolution
\begin{equation}
  \label{evol-spin}
  U(T) = e^{i\left(\sum_{i,j}\vec{\sigma}_i J_{ij} \vec{\sigma}_j +
      \sum_i\vec{B}_i\vec\sigma_i\right)T},
\end{equation}
where the matrices $J_{ij}$ and the vectors $\vec{B}_i$ determine the
spin Hamiltonian that we want to simulate. Since the real interaction
between the ions is described by more complicated Hamiltonians [See Eq.
(\ref{H-ions}) below], any of these transformations is an effective one,
realized after influencing the dynamics of the ions with external
fields.  This process, in which we dynamically modify the parameters of
the system ---Rabi frequencies, detunings, magnetic fields, etc---, in
order to achieve a well defined target operation, is called {\em
  Coherent Quantum Control}.

While coherent control has been implicit in any proposal for quantum computing
\cite{cirac95, poyatos98, sorensen99, cirac00, sorensen00, jonathan00,
  james00, milburn00, ripoll03, duan04, staanum04} and quantum simulation
\cite{porras04,barjaktarevic} with ion traps, the development of these schemes
of control relied always on the intuition of the researcher and on a clever
choice of approximations. In this work we show that many of these schemes can
be translated into a unified framework based on state dependent forces and
tunable traps. As characteristic examples, we will show how to implement the
pushing gate \cite{cirac00} and the fast gates based on laser pulses from
Ref.~\cite{ripoll03}. We will also propose an alternative and more general
model of fast gate based on continuous off-resonant lasers. As further
applications, we demonstrate how to induce collective $H=(\sum_i
\sigma^z_i)^2$ and nearest neighbor $H=\sum_i J\sigma^z_i\sigma^z_{i+1}$ Ising
interactions, and use them to produce squeezing, and generate cluster and GHZ
states of up to 30 atoms within a extremely short time, $T={\cal
  O}(1/\omega)$, where $\omega$ is the frequency of the ion trap.

Furthermore, within this unified framework we can address important
requirements of all these coherent processes. Namely they should: (i) be
independent of temperature (so that one does not need to cool the ions
to their ground state after they are moved to or from their storage
area); (ii) require no addressability (to allow the ions to be as close
as possible during the gate so as to strengthen their interaction), and
(iii) be fast (in order to minimize the effects of decoherence during
the gate, and to speed up the computation). All of these requirements
can be formulated as constraints of the control problem, and as we will
see below, they can be easily solved. Last but not least, we study the
scaling of resources as we try to make our control faster and answer the
question of what is the ultimate limit for the speed of our quantum
gates or entangling processes, which is shown to be determined both by
dissipation and non-harmonic contributions to the restoring forces.

The paper is organized as follows. In Sec. \ref{sec:theory} we develop the
formalism to study our system. First of all, we introduce the Hamiltonian for
any number of ions subject to state dependent forces and quasi-1D confinement,
and derive the harmonic approximation and the normal modes.  Then, we show how
to implement a unitary transformation, $U=\exp[i\phi(\{\sigma^z_i\})]$, made
of robust geometrical, while leaving the motional state unchanged.  In
Sec.~\ref{sec:fast-2qb} we apply our methods to quantum computing in two-qubit
setups. We demonstrate how to recover previous designs of a phase gate,
including the adiabatic pushing gate \cite{cirac00} and the fast gate based on
$\pi/2$ laser pulses that kick the ions \cite{ripoll03}.  Most important,
since the generation of perfect and very short laser pulses is a difficult
task, we design an arbitrarily fast quantum gate based on continuous laser
sources.  In Sec.~\ref{sec:control} we study the coherent control of many-ion
setups. We prove that arbitrary spin interactions of the type
(\ref{evol-spin}) can be simulated with the appropriate time-dependent forces,
and develop a numerical method to find them. As applications, we demonstrate
numerically the creation of squeezed, cluster and GHZ states of up to 30 ions
in a very short time.  Sec.~\ref{sec:errors} focuses on the study of errors.
First of all we introduce a model for dissipation on the vibrational degrees
of freedom. Next, we solve this model exactly and analyze the fidelity of the
effective unitary operations (\ref{phase-gate})-(\ref{evol-spin}). We prove
that for a perfect control and no dissipation our scheme is insensitive to
temperature.  Furthermore, the errors due to interaction with the environment
can be computed and optimized using the same tools of quantum control as in
Sec.~\ref{sec:fast-2qb}-\ref{sec:control}.  Finally, we study the errors due
to anharmonic terms in the interaction and show that both this and dissipation
set an upper bound on the speed of the gate. In Sec.~\ref{sec:conclusions} we
summarize our results and offer perspectives for future work.


\section{Theory}
\label{sec:theory}

\subsection{Normal modes}

\label{sec:normal-modes}

As mentioned before, this paper studies a set of $N$ ions, in an
essentially one-dimensional confinement\footnote{This is only for
  illustrative purposes. More complex setups, in which motion is allowed
  along all directions of space can be studied with exactly the same
  tools.} and subject to some external forces. Our model Hamiltonian is
\begin{equation}
  \label{H-ions}
  H = \sum_i \left[\frac{p_i^2}{2m}+V_{e,i}(x_i)-F_i(t)x_i\right]
  +\sum_{i<j} \frac{e^2}{4\pi\epsilon_0}\frac{1}{|x_i-x_j|}.
\end{equation}
In this equation, $V_{e,k}$ is the trapping potential that confines the
$k$-th ion, and it may be the same for all of them or may change from
ion to ion as in the case of microtraps \cite{cirac00}. The time
dependent external forces are denoted by $F_i(t)$ and, as we will see
below, they depend on the internal state of the ion, $\sigma^z_i$.

If we expand the previous energy around the equilibrium configuration without
forces, given by $(\partial H/\partial x_i)(x_1^{(0)},\ldots,x_N^{(0)}) = 0$
and $F_i=0$, we obtain a set of coupled harmonic oscillators
\begin{equation}
  \label{H-pushing}
  H = \sum_i \left[\frac{p_i^2}{2m}-F_i(t)x_i\right]
  +\frac{1}{2}m V_{ij}x_ix_j + E_0.
\end{equation}
The constant $E_0$ is the energy of the ions at their equilibrium
positions. The matrix $V$ describes the restoring forces: it is
symmetric, positive definite and it can decomposed as $V = M
\Omega^2 M^t$, with a positive diagonal matrix of frequencies,
$\Omega_{kl} = \omega_k \delta_{kl}$, and an orthogonal
transformation ($M M^t = M^t M = \mathbb{I}$). Using this
canonical transformation to define the normal modes $x_i = \sum_k M_{ik}
Q_k$, $p_i = \sum_k M_{ik} P_k$, we arrive at the Hamiltonian
\begin{equation}
  H = \sum_k \left[\frac{P_k^2}{2m} + \frac{1}{2}m\omega_k^2Q_k^2
    - \sum_i F_i M_{ik} Q_k\right].
\end{equation}
It is now useful to write this Hamiltonian in dimensionless form,
using the characteristic length of the oscillators, $\alpha_k =
(\hbar/m\omega_k)^{1/2}$, so that $P_k = \hbar \tilde
P_k/\alpha_k$ and $Q_k = \alpha_k \tilde Q_k$, $\tilde M_{ik} =
M_{ik}\alpha_k$. With this we arrive to
\begin{equation}
  H = \sum_k \left[
    \frac{1}{2}\hbar\omega_k \left(\tilde P_k^2 + \tilde Q_k^2\right)
    - \sum_i F_i \tilde M_{ik} \tilde Q_k\right],
\end{equation}
or, using of Fock operators, $a_k \equiv (\tilde Q_k + i\tilde
P_k)/\sqrt{2}$,
\begin{equation}
  \label{H-normal}
  H = \sum_k \left[
    \hbar \omega_k \left(a_k^\dagger a_k + \tfrac{1}{2}\right)
    - \sum_i \frac{F_i \tilde M_{ik}}{\sqrt{2}} (a_k^\dagger+a_k)\right].
\end{equation}


\subsection{Robust phases}

\label{sec:geom-phase}

We will now demonstrate how to obtain a robust phase by applying
forces to a harmonic oscillator. Let us begin with the toy model
\begin{equation}
  \label{HO-1}
  H = \hbar\omega a^\dagger a - f(t)\tfrac{1}{\sqrt{2}}(a^\dagger + a).
\end{equation}
When integrating the Schr\"odinger equation associated with this
Hamiltonian, it will convenient to use the overcomplete basis of
coherent states
\begin{equation}
  |z\rangle :=
  e^{-|z|^2/2}\sum_n \frac{z^n}{\sqrt{n!}}|n\rangle =:
  | Q+i P\rangle.
\end{equation}
Under the Hamiltonian (\ref{HO-1}), the coherent states behave somehow
like classical particles in phase space, because their center, given by
$\langle Q\rangle$ and $\langle P\rangle$, follows a classical
trajectory, while the width of these wavepackets, given by the
uncertainty of $Q$ and $P$, remains fixed.

\begin{figure}[t]
  \resizebox{\linewidth}{!}{\includegraphics{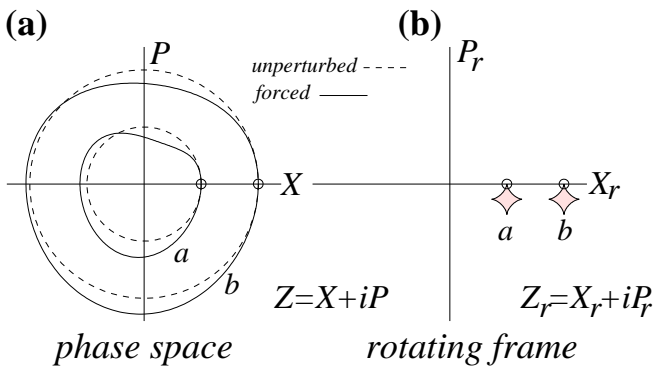}}
  \caption{ Trajectories on phase space of a coherent wavepacket subject
    to a single forced harmonic oscillator, $H=\omega a^\dagger a + F
    \sin(2t)(a+a^\dagger)$. In Fig. (a) we plot the usual phase space
    trajectories, $\langle a\rangle = \langle X+iP\rangle/\sqrt{2}$,
    without forcing (dashed) and with F=0.1 (solid), for two initial
    conditions. In Fig.~(b) we plot the same, but on a rotating frame of
    reference, $\langle a\rangle = \langle e^{-i\omega
    t}(X_r+iP_r)\rangle/\sqrt{2}$. }
  \label{fig:traj}
\end{figure}

More precisely, for an initial coherent state, $|\psi(0)\rangle =
|z_0\rangle$, the solution to the Schr\"odinger equation
$i\hbar\dot\psi = H(t) \psi$ is given also by a coherent state
$|\psi(t)\rangle = e^{i\phi(t)}|z(t)\rangle$, whose phase and
position satisfy
\begin{subequations}
\begin{eqnarray}
  \frac{dz}{dt} &=& -i \omega z + i \tfrac{1}{\sqrt{2}\hbar}f(t),\\
  \frac{d\phi}{dt} &=& \tfrac{1}{2\sqrt{2}\hbar}f(t)(\bar z + z).
\end{eqnarray}
\end{subequations}
The first equation has a solution,
\begin{equation}
  \label{position}
  z(t) = e^{-i\omega t}\left[z_0 +
    \tfrac{i}{\sqrt{2}\hbar}\int_0^td\tau\,e^{i\omega\tau}f(\tau)\right],
\end{equation}
that results from composing a displacement with a rotation of
angular speed $\omega$. Using the rotating phase-space
coordinates, $z_r := e^{i\omega t}z =: Q_r + iP_r$, we get rid of
the motion due to the unforced harmonic oscillators and find
\begin{subequations}
\begin{eqnarray}
  \frac{dz_r}{dt} &=& ie^{i\omega\tau}f(\tau)/\hbar\sqrt{2},\\
  \frac{d\phi}{dt} &=& \mathrm{Im}\,\frac{dz_r}{dt}\bar z_r
  = \frac{dP_r}{dt}Q_r -\frac{dQ_r}{dt}P_r
  = 2\frac{dA}{dt}.\label{phase-1}
\end{eqnarray}
\end{subequations}
The last equation means indeed that the growth of the phase is proportional to
the area spanned by the coherent state as it moves through the phase space.
The phase is not only geometrical in this sense, but also in the extended
definition of geometrical phase given in Ref. \cite{aharonov87}. Applying the
formulas in the previous reference one finds that if the total phase is
$\phi(t)$, and the dynamical phase is always twice the geometrical one,
$\phi_d(t)=-2\phi_g(t)$, so that in the end $\phi(t)=-\phi_g(t)$.

In this work we are interested in this phase and on making it
depend on the internal state of the particles governed by the
oscillator. We want, however, neither to influence the motional
state of the particles, nor to entangle internal and motional
degrees of freedom. For this reason we set a time limit on the
duration of the force and impose that after a fixed time $T$ the
coherent wavepacket is restored to its original state
\begin{equation}
  \label{condition-1}
  \int_0^T d\tau\, e^{i\omega \tau}f(\tau) = 0.
\end{equation}
Using this condition we derive a simple formula for the total phase
\begin{eqnarray}
  \label{phase}
  \phi(T) &=& \mathrm{Im}
  \int_0^T d\tau\,\tfrac{i}{\sqrt{2}\hbar}e^{i\omega\tau}f(\tau)
  \,\bar z_r(\tau)\\
  &=& \tfrac{1}{2\hbar^2}
  \mathrm{Im}\int_0^T d\tau_1 \int_0^{\tau_1}d\tau_2\,
  e^{i\omega(\tau_1-\tau_2)} f(\tau_1)f(\tau_2).\nonumber
\end{eqnarray}

As a simple application, in Fig.~\ref{fig:traj} we show the phase space
trajectories obtained by forcing two coherent states with a sinusoidal
force, $F(t)\propto \sin(2\omega t)$, where $\omega$ is the frequency of
the Harmonic oscillator. Even though looking at Fig.~\ref{fig:traj}(a)
the orbits of different initial conditions seem also very different, on
the rotating frame of reference the enclosed area $A$ is always the same
[Fig.~\ref{fig:traj}(b)]. In other words, the phase $\phi$ is
insensitive to the initial motional state of the system and it is thus
robust. This property is of crucial importance when we seek applications
to real systems that cannot be cooled to the zero phonon limit, but
which thanks to Eq.~(\ref{condition-1}) will pick up the same phase
regardless of the temperature.


\subsection{Phase of two ions}
\label{sec:geom-phase-2}

We will now apply the results from Sec. \ref{sec:geom-phase} to a pair
of ions. In this case there are two normal modes: the center of mass,
$x_c=(x_1+x_2)/2$, which oscillates with frequency $\omega_c$, and the
stretch mode, $x_s = x_2-x_1$, which oscillates with frequency
$\omega_s$. If the ions are stored in the same harmonic trap,
$V_{e,k}(x_k) = \frac{1}{2}m\omega^2x_k^2$, these frequencies are found
to be incommensurate, $\omega_c=\omega$ and $\omega_r=\omega\sqrt{3}$.
If we store the ions in two microtraps (or in a more complicated
potential), the value of $\omega_r$ can be tuned from $\omega\sqrt{3}$
down to $\omega$.

If we exert a similar state dependent force on both ions, for instance
by means of an off-resonance laser that induces a AC Stark shift on one
of the internal state of the ions, the Hamiltonian (\ref{H-ions}) will
look as follows
\begin{eqnarray}
  H &=& \hbar\omega a_c^\dagger a + \hbar\omega_r a_r^\dagger a_r
  -F(t)\sigma^z_1 x_1 - F(t)\sigma^z_2x_2\\
  &=& \hbar\omega a_c^\dagger a + \hbar\omega_r a_r^\dagger a_r
  + F(t)(\sigma^z_2-\sigma^z_1)d\nonumber\\
  &-& F(t)(\sigma^z_1+\sigma^z_2)\frac{\alpha_c}{\sqrt{2}}
  (a_c+a_c^\dagger)\nonumber\\
  &-& F(t)(\sigma^z_2-\sigma^z_1)\frac{\alpha_r}{2\sqrt{2}}
  (a_r+a_r^\dagger)\nonumber
\end{eqnarray}
where $d$ is the equilibrium distance between the ions,
$\alpha_{c,r}^2=\hbar/m\omega_{c,r}$ are the lengths of the oscillators,
and $\sigma^z_i$ is an operator that has value $+1$ or $-1$ depending on
whether the ion is on internal state $|+1\rangle$ or $|-1\rangle$.

Since the modes are now decoupled, we can apply the formulas of
Sec.~\ref{sec:geom-phase} almost directly. We first obtain a pair
of commensurability relations on the force
\begin{equation}
  \label{condition-2}
  \int_0^T d\tau\, e^{i\omega_{c,r}\tau}F(\tau) = 0,
\end{equation}
which are just a generalization of Eq.~(\ref{condition-1}). Next we obtain the
total phase, which up to local and global contributions is
\begin{equation}
  \label{phase-2}
  \phi = \sigma^z_1\sigma^z_2
  \int_0^Td\tau_1\int_0^Td\tau_2\,{\cal G}(\tau_1-\tau_2) F(\tau_1)F(\tau_2)
\end{equation}
where ${\cal G}(t) = \left[\frac{1}{\omega_c}\sin(\omega_c |t|) -
\frac{1}{\omega_r}\sin(\omega_r |t|)\right]/2m\hbar$.

The goal in Sec.~\ref{sec:fast-2qb} will be to tune the forces so
that Eq.~(\ref{condition-2}) is satisfied and the phase becomes
$\phi = \pi\sigma^z_1\sigma^z_2/4$, the required value for a
two-qubit quantum phase (\ref{phase-gate}). We will then show two
optimal solutions to this problem, which use either pulsed or
continous forces.


\subsection{Phase for any number of ions}
\label{sec:geom-phase-N}

The case of $N>2$ ions exhibits a richer structure, due to the fact that
the phase depends on all pair products $\sigma^z_i\sigma^z_j$ of the
polarizations of the atoms. If we apply the formula for the phase of one
harmonic oscillator (\ref{phase}) to each of the modes in the effective
Hamiltonian for the chain (\ref{H-normal}), we obtain a total phase
\begin{equation}
  \label{proto-ising}
  \phi = \sum_{ij} \sigma^z_i\sigma^z_j \int_0^T\int_0^Td\tau_1d\tau_2\,
  F_i(\tau_1) {\cal G}_{ij}(\tau_1-\tau_2) F_j(\tau_2),
\end{equation}
with a Hermitian kernel
\begin{equation}
  G_{ij}(t) = \sum_k \frac{M_{ik}M_{jk}}{2m\omega_k\hbar}
  \sin(\omega_k|t|),
\end{equation}
plus a generalization of Eq. (\ref{condition-2})
\begin{equation}
  \label{condition-N}
  \sum_i \int_0^T d\tau e^{i\omega_k\tau}\alpha_kM_{ik}F_i(\tau) =
  0,\quad\forall k.
\end{equation}
In Sec.~\ref{sec:control} we will show how Eq.~(\ref{proto-ising})
can be related to an effective Ising interaction, $H=\sum_{ij}
J_{ij}\sigma^z_i\sigma^z_j$, whose precise shape can be engineered
and which can produce interesting entangled states.


\section{Fast phase gate for two ions}
\label{sec:fast-2qb}

A very important application of the techniques studied so far is to
design a two-qubit quantum gate that is robust enough to be included in
a scalable ion-trap quantum computer. This task has been pursued in a
previous work \cite{ripoll03} using a slightly more complicated method,
in which the force was achieved by kicking the ion with $\pi/2$ laser
pulses, and the distribution and number of these pulses had to be
designed ``by hand''. In this section we review this work in the light
of our new formalism and rephrase it as an optimal control problem. This
allows us to consider more general forces, and to find, for instance, a
design of a gate that involves the shortest time, and the weakest and
smoothest varying forces.

\subsection{Kicking forces}

We will consider two ions in a one-dimensional harmonic trap,
interacting with a laser beam on resonance. The Hamiltonian
modeling this system can be written as $H=H_0 + H_1$ where $H_0 =
\hbar \omega_c a^\dagger_c a_c + \hbar\omega_r a_r^\dagger a_r$
describes the normal modes of the ions and
\begin{eqnarray}
  \label{H-kick}
  H_l &=& \frac{\Omega(t)}{2}\left[ \sigma^+_1 e^{i\hbar k x_1}
    +\sigma^+_2 e^{i\hbar k x_2} + \mathrm{H. c.}\right]\\
  &=& \frac{\Omega(t)}{2}\left[\sigma^x_1e^{-i\hbar k x_1\sigma^z_1}
   + \sigma^x_2e^{-i\hbar k x_2\sigma^z_2}\right].
\end{eqnarray}
This term describes processes in which the internal state of an ion is
changed and, as a consequence of the absorption and emission of photons,
the atom gains momentum, $\hbar k$. The Rabi frequency $\Omega(t)$ is a
function of the intensity of the lasers that induce these internal
transitions, and looking for the simplest setup we assume that it is the
same for both ions.

In Ref.~\cite{ripoll03} we explained how to use the Hamiltonian
$H_l$ to kick the ions. The process consists on applying very fast
laser pulses, in which the Rabi frequency, $\Omega(t)$, and the
duration of the pulse, $\delta t$, satisfy $\int_0^{\delta
t}\Omega(\tau)d\tau = \pi$ and $\delta t \ll 2\pi/\omega$. Let us
study the evolution of the ions under a single laser pulse. Since
the pulse is much shorter than a period of the trap, we can
neglect the influence of $H_0$.  We then use the formula
\begin{equation}
{\cal T}\exp\left[i\int_0^{\delta
t}d\tau\,\frac{\Omega(\tau)}{2}\;\vec{n}\vec{\sigma}\right] =
\cos(\theta) + i \sin(\theta) \vec{n}\vec{\sigma},
\end{equation}
where $\Vert \vec n\Vert=1$ is a unitary vector and we impose that a
$\pi/2$ pulse is produced: $\theta = \int_0^{\delta t}\Omega(\tau)/2 =
\pi/2$. Under these conditions the unitary evolution is described by
\begin{equation}
  U_{kick} = \sigma^x_1\sigma^x_2
  e^{i\hbar k(x_1\sigma^z_1+x_2\sigma^z_2)}.
\end{equation}
If at times $\{t_1,t_2,\ldots,t_{N_p}\}$ we send groups of
$\{2n_1,2n_2,\ldots,2n_{N_p}\}$ very short laser pulses with alternating
momenta, $+k,-k,+k,\ldots$, the total operation can be written as $U =
\prod_l \exp[n_l\times i\hbar k(x_1\sigma^z_1+x_2\sigma^z_2)]$, and can
be thought of as induced by the effective force
\begin{equation}
  \label{kicking-force}
  F_i(t) = \sum_{l=1}^{N_p} 2n_l\times \hbar k \sigma^z_i \delta(t-t_{l}).
\end{equation}
When the number of pulses is odd, similar considerations can be done,
but now a total spin flip $\sigma^x_1\sigma^x_2$ has to be included by
hand at the end of the process.


\subsection{Phase gate based on kicks}
\label{sec:scheme-kicks}

The parametrization of Eq.~(\ref{kicking-force}) is a means to simplify
the problem of finding the optimal forces that produce the phase gate
(\ref{phase-gate}). Using the previous notation, the conditions for
restoring the motional state of the ions become
\begin{equation}
  \label{condition-kicks}
  \sum_{l=1}^{N_p} n_l e^{-i \omega_{c,r} t_k}=0.
\end{equation}
If these equations are satisfied, the accumulated phase will be
\begin{equation}
  \phi = \sum_{l,m=1}^{N_p}
  \left[\frac{\sin(\omega_c t_{lm})}{\omega_c}
    - \frac{\sin(\omega_r t_{lm})}{\omega_r}\right]
  \frac{2\hbar k^2 n_l n_m}{m} = \pi/4,
\end{equation}
where $t_{lm} = |t_l-t_m|$ is the time between the $l$-th and the
$m$-th kicks.

\begin{figure}[t]
\centering
\resizebox{\linewidth}{!}{\includegraphics{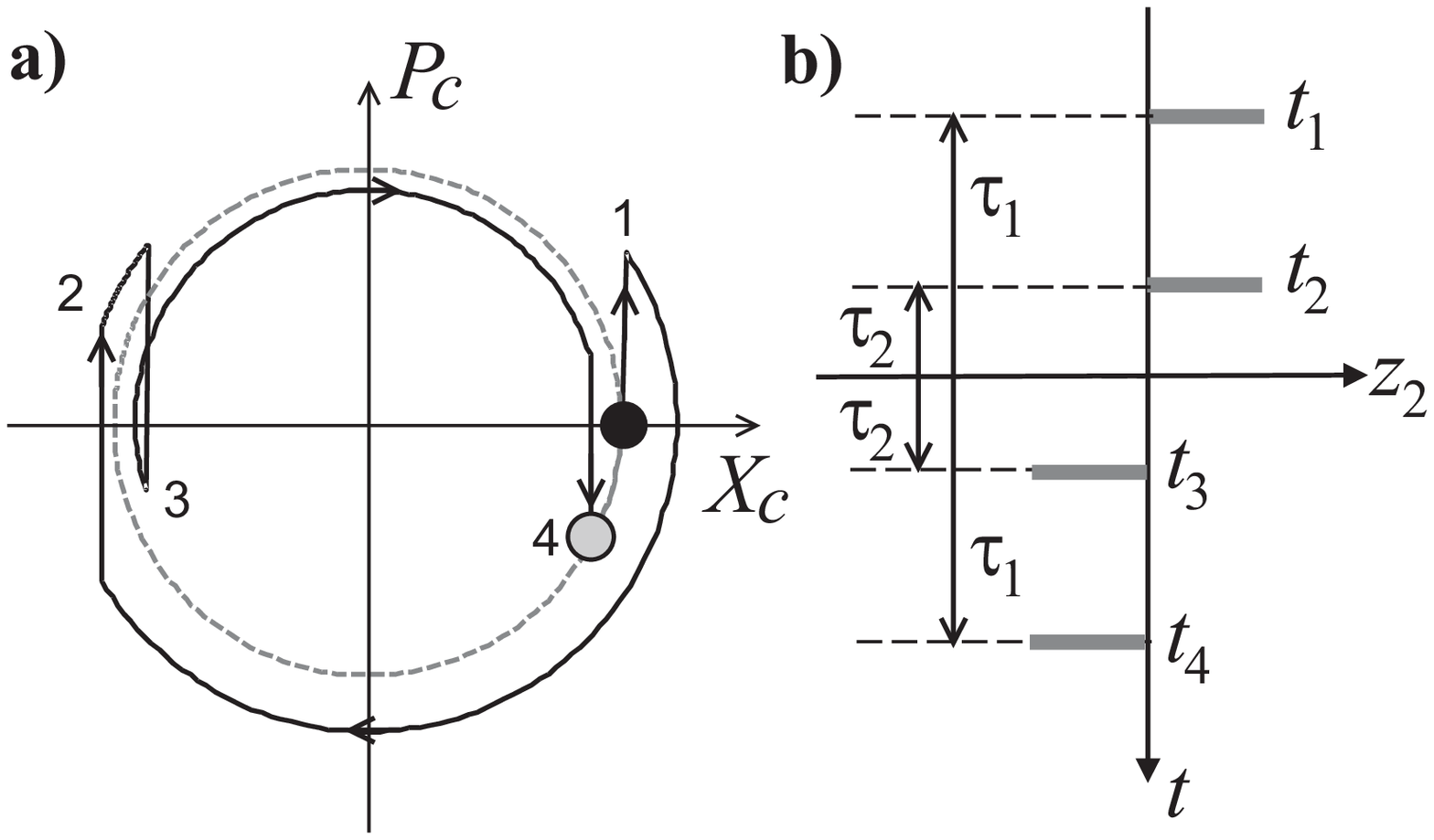}}
\resizebox{\linewidth}{!}{\includegraphics{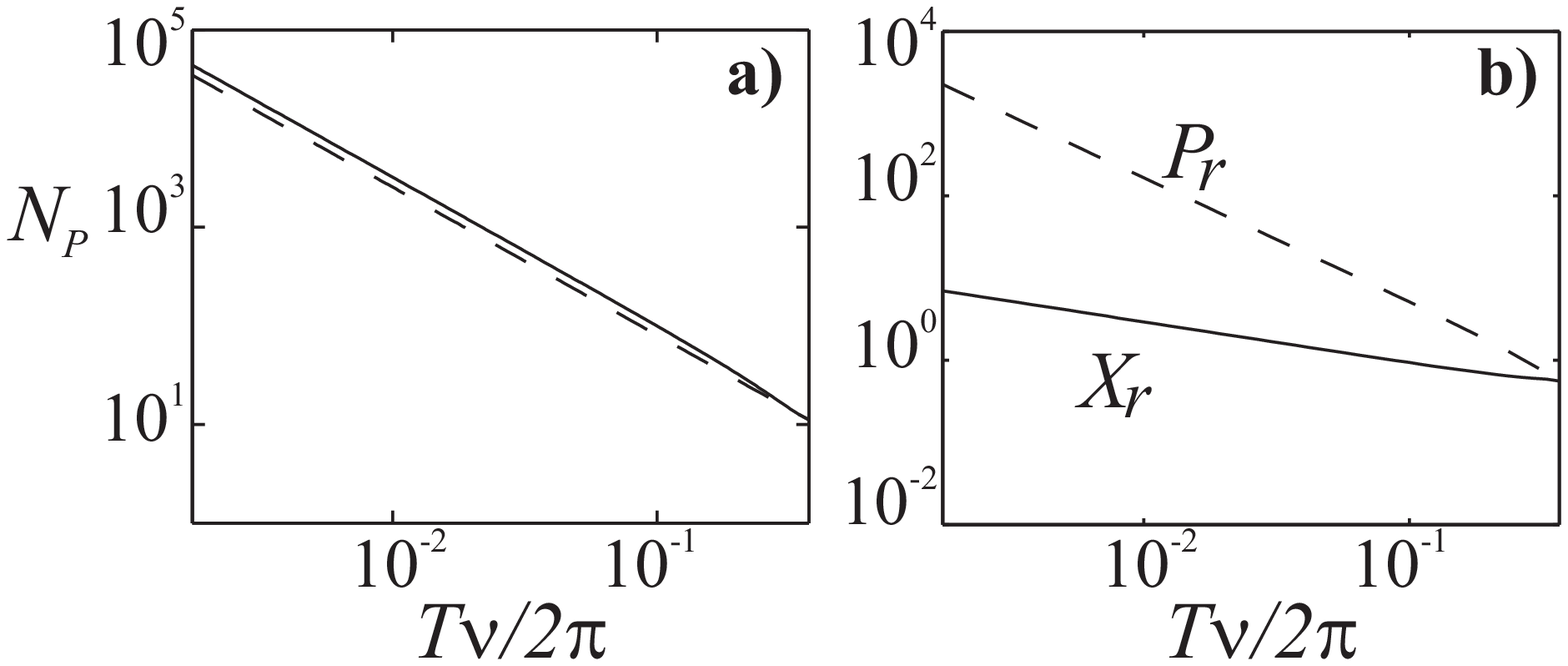}}
\caption{(a) Trajectory in phase space of the
  center-of-mass state of the ions $(\tilde Q_c,\tilde P_c)$ during the
  2-qubit gate (solid line), connecting a possible initial state (black
  filled circle) to its final state (grey filled circle), for the
  protocol designed in Sec.~\ref{sec:scheme-kicks}. Figure (b) shows how
  the laser pulses (bars) distribute in time for this scheme. Below we
  plot (c) the Number of pairs of pulses, and (b) relative angle of the
  two laser beams required to produce a phase gate using this scheme.}
\label{fig:kicks}
\end{figure}

In Ref.~\cite{ripoll03} we have found two possible solutions for
these equations. The first protocol that we proposed, performs the
phase gate in a time $T=1.08/\nu$, using about 4 pulses, while the
second protocol allows for an arbitrarily short gating time $T\sim
N_p^{-2/3}/\nu$ at the expense of using more pulses or kicks.

The method for the first protocol is illustrated in
Fig.~\ref{fig:kicks}(a-b), where we plot the phase space trajectories
followed by the center of mass mode. This sequence consists basically on
four groups of pulses given by $(n_l/n, t_l)= \{(\gamma,-\tau_1),
(1,-\tau_2), (-1,\tau_2), (-\gamma,\tau_1)\}$. The number $n$ tells us
how many pulses are sent within each group, and the parameter $0 <
\gamma = \cos(\theta) < 1.0$ is a real number that describes how much
tilted are the kicking lasers with respect to the axis of the trap. It
is always possible to find a solution to Eq.~(\ref{condition-kicks})
with $\tau_1 \simeq 0.538(4)(2\pi/\nu) > \tau_2 > 0$.  The results for
the performance of the gate are summarized in Fig.~\ref{fig:kicks}(c-d):
for realistic values of the Lamb--Dicke parameter \cite{demarco02} we
only need to apply the sequence of pulses one or two times to implement
a phase gate.

The second protocol performs the gate in an arbitrarily short time $T$.
For shortening the time we now require six groups of pulses, distributed
according to $(n_l/n,t_n)= \{(-2,-\tau_1), (3,-\tau_2), (-2,-\tau_3),
(2,\tau_3), (-3,\tau_2), (2,\tau_1)\}$, where the times $\tau_1$,
$\tau_2$ and $\tau_3$ are found numerically by solving the
transcendental equations (\ref{condition-kicks}), with the
constraint that the whole process takes a time $T=2\tau_1$. As
Fig.~\ref{fig:kicks} shows, the number of pulses, $N=14n$, increases
with decreasing time as $N_p\propto T^{-3/2}$. This is just a
consequence of a more general result that is shown later.


\subsection{Phase gate based on continuous forces}
\label{sec:continous}

The use of $\pi/2$ pulses to introduce momentum in the ions has
some inconveniences. First of all, each of the pulses has to be
perfect, and induce a complete population transfer from one
internal state to the other one. If this is not the case,
systematic errors on each of the pulses can lead to an exponential
decrease of the fidelity. Furthermore, as we increase the gating
speed, the pulses may become too long to be considered as
instantaneous kicks, and the previous formalism fails.

What we have found, and what is also one of the main results of this
paper, is that the phase gate may be produced also by applying
continuous forces. The search for this forces is then no more difficult
than solving an eigenvalue equation, where one may add restrictions such
as smoothness of the force, and minimal total work.

Let us take the real vector space $L^2([0,T])$ of space of square integrable
real forces in the $[0,T]$ interval, with the usual scalar product $(f,g) =
\int_0^t f(t)g(t)dt$. From this Hilbert space, we choose a subspace ${\cal
  H}$ of functions which are orthogonal to the modes $e^{i\omega_{c,r}t}$
\begin{equation}
  \int_0^T d\tau\,e^{i\omega_c\tau}F(\tau) =
  \int_0^T d\tau\,e^{i\omega_r\tau}F(\tau) = 0.
\end{equation}
Within ${\cal H}$, the phase and the smoothness of the gate are given by
$\phi[F]=(F,{\cal G} F)$ and $S[F]:=(F,-\frac{d^2}{dt^2}F)$, respectively. We
will prove that \textit{the optimal (i.~e.  smoothest) force that produces a
  phase gate $\phi=\pi/4$, is simply proportional to $F_\mu$, where $F_\mu$ is
  the eigenstate
\begin{equation}
  \label{eigenvalue}
  -\mu\frac{d^2}{dt^2}F_\mu = {\cal G} F_\mu
\end{equation}
with largest eigenvalue $|\mu|$. If rather than measuring the optimality with
$S[F]$ we use just the norm, $N[F] := \Vert F\Vert_2 = (F, F)$, then the
eigenvalue problem is simpler}
\begin{equation}
  \mu F_\mu = {\cal G} F_\mu.
\end{equation}

Let us prove this useful result. We have to work with four functionals, which
are the $S[F]$ and $\phi[F]$ defined above plus two more, which measure the
displacements originated by the force: $D_{r,c}[F] = \int_0^T d\tau\,
e^{i\omega_{c,r}\tau} F(\tau)$.  By choosing the space of real periodic
functions which are orthogonal to the Fourier modes $e^{i\omega_{c,r}\tau}$ we
ensure that everything is well defined and also that the constraints
$D_{c,r}[F] = 0$ are satisfied. This leaves us with the problem of finding a
force which minimizes $S[F]$, while satisfying the last constraint
$\phi[F]=\pi/4$.  There exists however a much easier dual problem which is
formulated as finding the maximum of $\phi[F]$ subject to the quadratic
constraint $S[F]=1$. Using the theory of Lagrange multipliers, this amounts to
finding the maximum of
\begin{equation}
  {\cal L}[F] = \phi[F] - \mu (S[F] - 1),
\end{equation}
where $\mu$ is the Lagrange multiplier. Differentiating the Lagrangian
we obtain Eq. (\ref{eigenvalue}), from where it follows that $\mu \Vert
F\Vert_2^2 = \phi$ is the maximal phase to be achieved, and the associated
eigenstate $F_\mu\sqrt{\pi/(4\mu)}/\Vert F\Vert_2$ is the force we were
looking for.

\begin{figure}[t]
  \centering
  \resizebox{\linewidth}{!}{\includegraphics{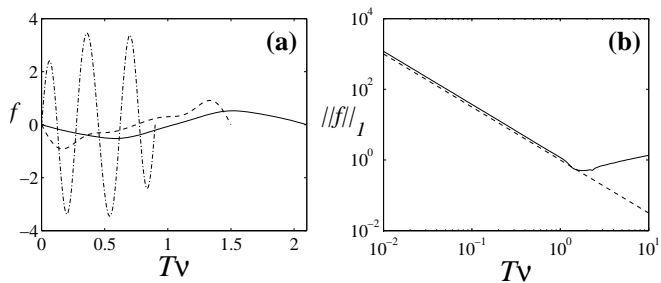}}
  \caption{(a) Optimal forcing for a gating time $T\nu=2.1$ (solid),
    $1.5$ (dashed) and $0.9$ (dash-dot), where $\omega = 2\pi\times\nu$
    is the frequency of the ion trap. (b) Intensity of the force vs.
    duration of the gate (solid) and visual aid $(T\nu)^{-3/2}$
    (dashed). All magnitudes have been adimensionalized following the text.}
  \label{fig:2qb}
\end{figure}

Even though we have been able to relate the control problem to an eigenvalue
equation (\ref{eigenvalue}), there exists no simple analytical solution to
this problem and we have to resort to some simple numerics. However, a very
nice feature of the two-ion problem is that, by scaling quantities with
respect to the trap strength, $\omega$, and the wavepacket size, $\alpha =
(\hbar/m\omega)^{-1/2}$, we can compute the optimal force independent of the
setup. Using these units and expanding the force in term of Fourier modes
\begin{equation}
  f(t) = \sum_{m=-N_m}^{N_m} c_n e^{i 2m\pi t/T} =
  \frac{F(t)\alpha}{\hbar \omega},
\end{equation}
we can express Eq.~(\ref{eigenvalue}) as an eigenvalue equation
for the vector $\vec{c}$, that is to be solved numerically. The
number of modes $N_m$ can in principle be any number above 3,
because some degrees of freedom are lost when satisfying the
constraint (\ref{condition-2}). However, the numerical experiments
show that indeed $N_m=4$ provides very good solutions. As an
example, in Fig. \ref{fig:2qb}(a) we show three possible forces
for a duration of the gate $T\omega/2\pi= 2.1, 1.5$ and $0.9$. We
have computed other solutions for a wider range of gate speeds. In
Fig.~\ref{fig:2qb}(b) we plot the mean intensity $\Vert
f\Vert=\int_0^T|f(t)|dt$ versus the total time $T$, and
demonstrate the law $T^{-3/2}$ obtained above.

An interesting question is how much energy we have to put into the
system in order to produce faster and faster gates. With our current
formalism, this can be answered very quickly. Let us assume that we have
the optimal force that produces a phase gate in time $T\ll
2\pi/\omega_{c,r}$. Since the time is very short, we can perform a
Taylor expansion of the function ${\cal G}(\tau)$ obtaining
\begin{eqnarray}
  |\phi| &\simeq& \left|\int_0^T\int_0^Td\tau_1d\tau_2 F(\tau_1)F(\tau_2)
  \frac{(\omega_c^2-\omega_r^2)}{12m\hbar}|\tau_1-\tau_2|^3\right|\nonumber\\
  &\leq& \frac{\omega_c^2-\omega_r^2}{12m\hbar} T^3 \Vert F\Vert_1^2
\end{eqnarray}
where $\Vert F\Vert_1 = \int_0^T d\tau\, |F(\tau)|$ is just a measure of
the force applied. From here we see that
\begin{equation}
  \Vert F\Vert_1 \simeq T^{-3/2}
\end{equation}
or, in the case of the kicked ions, $N_p \simeq T^{-3/2}$, the scaling
that the numerical simulations already showed.


\subsection{Adiabatic pushing}

As a final remark, we want to relate the methods presented here with the
\emph{pushing gate} introduced in Ref. \cite{cirac00}. That work proposed to
trap the ions in separate microtraps, $V_{e,i}(x_i) =
\frac{1}{2}m\omega^2(x_i-x_i^{(0)})^2$, and to apply a state dependent
force on two neighboring ions. This force should be switched on and off
adiabatically with respect to the period of the traps, $2\pi/\omega$, in
order to approach the ions to each other and later on return them to
their equilibrium positions. The adiabaticity condition ensures that the
ions remain at all times in the ground state of the effective
Hamiltonian (\ref{H-pushing}), which is now time dependent, because the
equilibrium positions $x_k^{(0)}$ and the equilibrium energy $E_0$ both
depend on the instantaneous value of the forces. After restoring the
ions to their original positions, the only effect on the quantum state
of the ions is a state-dependent phase
\begin{equation}
  \phi = \int_0^T E_0\left[x_1^{(0)}(t),\ldots,x_N^{(0)}(t)\right] dt.
\end{equation}
The previous analysis is found in Refs. \cite{cirac00,calarco01,sasura03}. A
very important point is that, in order not to excite the ions and regard the
process as truly adiabatic, the forces have to be weak and change very slowly,
and no cubic contributions to the energy should appear. In other words, we
should be able to describe the change of $E_0$ using at most quadratic terms,
$E_0[\vec{x}^{(0)}(t)] \simeq E_0[\vec{x}^{(0)}(0)] + \frac{1}{2}mV_{ij}\delta
x_i^{(0)}\delta x_j^{(0)}$. Hence, rather than using the adiabatic theorem we
can integrate the problem exactly. For a single harmonic oscillator we get
\begin{equation}
  z(t) = e^{-i\omega t}z_0 + \frac{1}{\sqrt{2}}f(t)
  - \frac{1}{\sqrt{2}}\int_0^td\tau\,e^{i\omega\tau}\frac{1}{\omega}
  f'(\tau),
\end{equation}
where the adiabatic condition corresponds to neglecting the last term
$f'(\tau)/\omega$, and the force only has to satisfy
$f(T)=f(0)=0$. Repeating the arguments of previous sections, for two
ions in neighboring traps the total phase becomes
\begin{equation}
  \phi = \int_0^T d\tau
  \frac{\omega_c^{-1}-\omega_r^{-1}}{2m\hbar}F(\tau)^2
  \sigma^z_1\sigma^z_2.
\end{equation}
Here $\omega_c=\omega$ and $\omega_r$ now depend slightly on the
separation of the microtraps, but the same formula applies for the case
in which both ions coexist in the same trap ---a situation that could
not be considered with the formalism of previous papers.


\section{Quantum control of several ions}
\label{sec:control}

We will now study 1D setups with more than two ions. As we showed before, we
can still control the geometric phases and use them to simulate a variety of
spin Hamiltonians [Sec.~\ref{sec:simulation}]. The design of the forces for
these simulations is still a control problem, but a much more difficult one.
For instance, a crucial difference is that in setups with more than two ions
either addressability or a spatial modulation of the forces are required. As a
possible application of this result we study how to optimally generate
entangled states and squeezing. In particular, we show that this can be done
for a large number of ions (up to 30) in a very short time
[Sec.~\ref{sec:entangler}].

\subsection{Simulation of spin Hamiltonians}
\label{sec:simulation}

\textit{Given any Ising Hamiltonian
\[H_J = \sum_{ij} J_{ij}\sigma^z_i\sigma^z_j,\quad
J=J^t \in \mathbb{R}^{N\times N},\] and a time, $T$, it is possible to
design a set of state dependent forces, $F_i(t;J)$ such that after
applying these forces for a time $T$, the dynamics of the ions simulates
this spin Hamiltonian. In other words
\[{\cal T}\left[e^{i \int_0^t H(\tau)d\tau}\right]
= e^{-iH_J T},\] where ${\cal T}$ denotes time ordered product and $H$ is the
true Hamiltonian of the ions (\ref{H-pushing}).}

The proof is very simple. Let us slice the time interval $[0,T]$
into $N^2$ subintervals, $0<t_{11}=t_{12}<\ldots<t_{N,N-1}<T$. In
a given time interval, $I=[t_{ij},t_{i,j+1}]$, we will switch on
two forces, and leave all other ions on their own,
\[F_k(t) = 0,\;t\in[t_{ij},t_{i,j+1}],\;\forall k\neq i,j.\]
The active forces $F_i(t)$ and $F_j(t)$ must satisfy several equations
\begin{eqnarray}
  &&0 =\int_I e^{i\omega_k \tau} M_{k\alpha}
  F_\alpha(\tau)d\tau,\; \alpha=i,j,\; k=1\ldots N\\ &&J_{ij}
  = \tfrac{1}{t_{i,j+1}-t_{ij}} \int_I
  \int_Id\tau_1d\tau_2\, F_i(\tau_1){\cal
  G}_{ij}(\tau_1-\tau_2) F_j(\tau_2).\nonumber
\end{eqnarray}
It is not difficult to convince oneself that these equations always have
a solution, and that by repeating this procedure we will get an
effective total phase, $\phi$, that resembles the one produced by the
Ising model during a time $T$. $\square$

We have to make several remarks here. The first one is that since the
operator that we want to simulate is symmetric, $J_{ij}=J_{ji}$, and since the
diagonal terms only contribute to a global phase, the number of
intervals can be actually decreased to $N(N-1)/2$.

However, more important is the fact that we can use coherent control to
find optimal forces, $\vec{F}$, which instead of piecewise
continuous are the smoothest possible and have the optimal norm, while
giving rise to the same effective Hamiltonian. This task has been
performed numerically for some models, and the results will be shown in
the following section.

From the point of view of quantum simulation, we would like to be able
to model more than just an Ising model, which is essentially classical.
For instance, one would like to be able to introduce transverse magnetic
fields, $\sum_i h_i\sigma^x_i$, or to simulate a Heisenberg interaction,
$\vec{\sigma}_i\vec{\sigma}_j$, and in general, a unitary operation of
the form (\ref{evol-spin}) would be desirable. The answer to this
problem is once more the stroboscopic evolution, or a Trotter expansion
of the operator (\ref{evol-spin}),
\begin{equation}
  U \simeq \left\{\prod_{\alpha=x,y,z}
  e^{i\frac{T}{N}
  t\left(\sum_{ij}J^\alpha_{ij}\sigma^\alpha_i\sigma^\alpha_j + \sum_i
  h^\alpha_i\sigma^\alpha_i\right)}\right\}^N.
\end{equation}
In this expansion, we decompose the total unitary as a product of phase
gates, that are originated by forces that depend on $\sigma^x_i$,
$\sigma^y_i$ and $\sigma^z_i$. In practice, one would switch on a
magnetic field $h^z_i$ and perform a phase gate with coefficients
$J^z_{ij}$ for a time $T/N$, rotate the spins so that $\sigma_y$ becomes
$\sigma_z$, apply the phases with $J^y_{ij}$, etc.

It is also worth noticing that if we switch on the state-dependent
forces acting on different ions, and make them oscillate with a single
frequency $\Omega$ around a constant value, $F_i(t) = f_i \sin(\Omega t)
\sigma^z_i$, for a long time, the effective interaction is a particular
spin Hamiltonian
\begin{equation}
  H  = \sum_{ijk} f_i
  \frac{M_{ik}M_{jk}[1+\delta_{\Omega,0}]}{4m\hbar (\omega_k^2-\Omega^2)} f_j
  \sigma^z_i\sigma^z_j + {\cal O}({f_i^2 \over \omega_k^2}),
\end{equation}
In the limit $\Omega\to0$, this model corresponds to the one found in
\cite{porras04}. As it was shown there, depending on whether the forces
operate longitudinally or transversely to the ion trap, this continuous
force will give rise to long range or short range interactions.


\subsection{Coherent control and design of entanglers}
\label{sec:entangler}

The simulation of an Ising interaction is by itself interesting, and has
important applications such as creation of many-qubit quantum gates, quantum
simulation and adiabatic quantum computing. However, a most straightforward
and useful application of an Ising Hamiltonian is the generation of many-body
highly entangled states called \emph{graph states} \cite{hein04}. Roughly
speaking, let us imagine that we have a set of $N$ spins, which we represent
by points or vertices, and let us connect these points by lines or edges. The
resulting graph can be described by an adjacency matrix which takes value
$J_{ij}=1$ only if the spins $i$ and $j$ are connected. To each graph we can
thus associate a Hamiltonian of the form $H_J = \sum_{i,j} J_{ij}
\sigma^z_i\sigma^z_j$. It has been shown that after applying this interaction
over a certain time on a transversely polarized state, the outcome is a highly
entangled state called a graph state:
\begin{equation}
  \label{graph-state}
  \left|\psi_G\right\rangle = \frac{1}{2^{N/2}}
  e^{-i\frac{\pi}8 H_{\cal G}}\left(\left|0\right\rangle
  +\left|1\right\rangle\right)^{\otimes N}.
\end{equation}
When the graph has a lattice geometry, these states are also known as
cluster states \cite{briegel01}, and form the basic ingredient of the
one-way quantum computer \cite{raussendorf01}. However, a particularly
important case without lattice geometry is the GHZ state,
\begin{equation}
  \left|\mathrm{GHZ}\right\rangle \sim \left|0\right\rangle^{\otimes N}
  + \left|1\right\rangle^{\otimes N},
\end{equation}
which is essentially generated by the interaction $J_{ij}=1$ or $H_J =
(\sum_i\sigma^z_i)^2$. The GHZ state is one of the best studied
entangled states, it constitutes a canonical example of Schr\"odinger
cat state, and it could have important applications in the field of
precision frequency measurements, providing a $1/\sqrt{N}$ precision
increase for $N$ entangled ions \cite{wineland92,wineland94}, a point
already demonstrated experimentally in Ref. \cite{meyer01}.

\begin{figure}[t]
  \centering
  \resizebox{\linewidth}{!}{\includegraphics{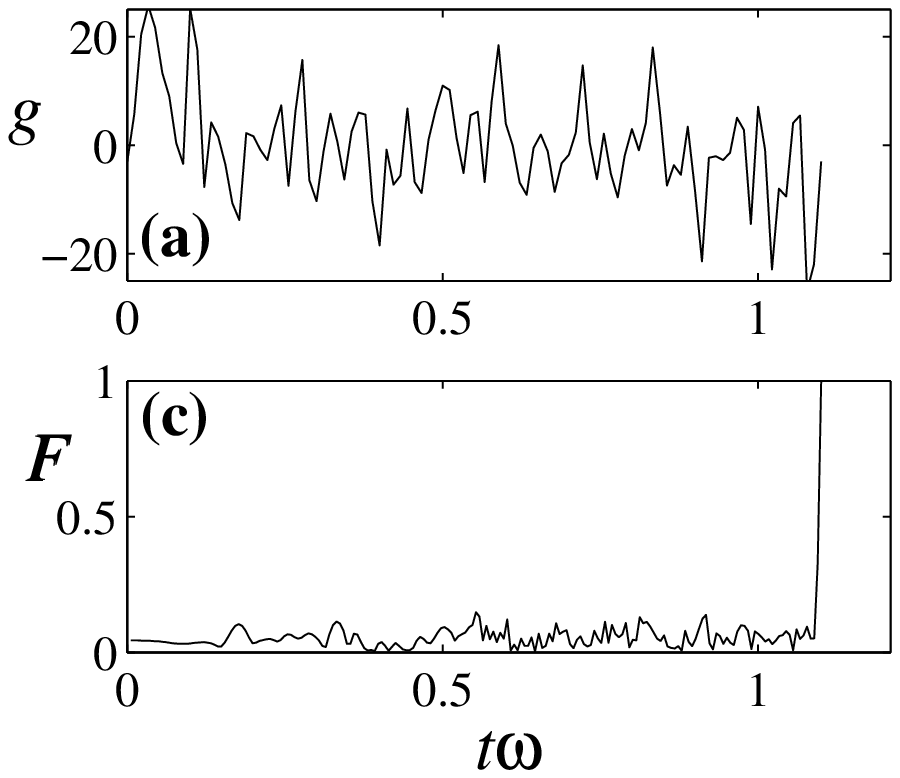}
    \includegraphics{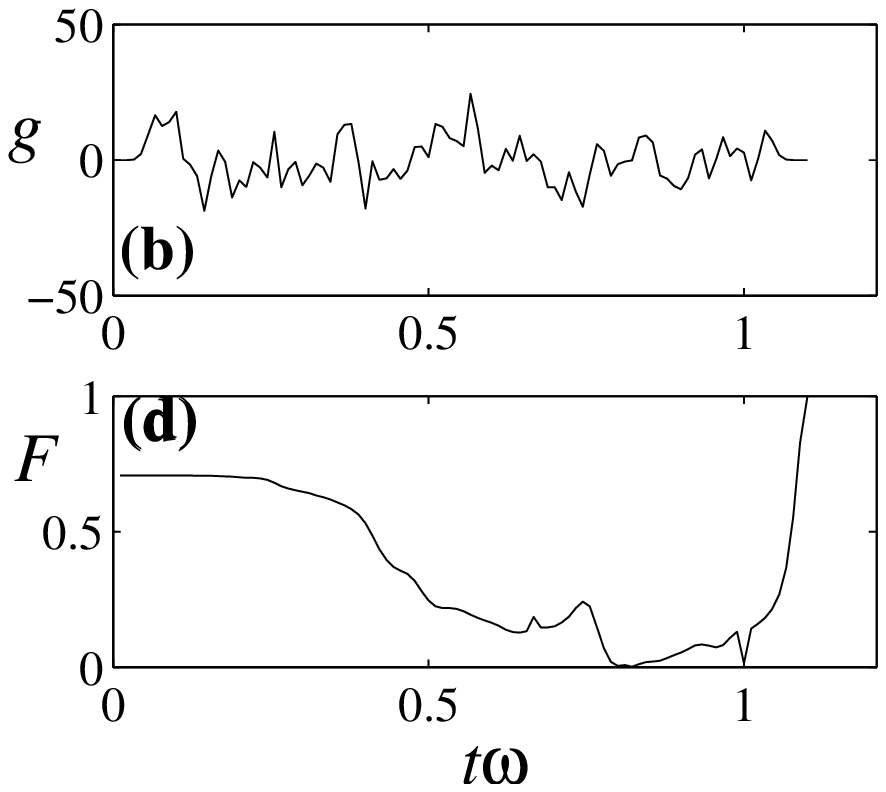}}
  \caption{
    Generation of a (a,c) cluster state of $N=10$ ions and of (b,d) a
    GHZ state of $N=20$ ions, using common forcing for a time
    $T=1.1/\omega$. (a,c) Time dependence of the forces, $f(t)$. (b,d)
    Fidelity with respect to the target state as a function of time.}
  \label{fig:entangling}
\end{figure}

We have investigated how to implement these highly entangled
states using our quantum control techniques. The idea is very
simple: we design a matrix $J_{ij}$ for our graph state, and look
for the time- and state-dependent forces that implement the phase
transformation $\exp(-i\pi H_J\pi/4)$ within a fixed time $T$. For
simplicity, even though it is not warranted to succeed, we look
for forces that have a common time dependence $F_i(t) = \gamma_i
f(t) \sigma^z_i$, $|\gamma_i| \leq 1$. These forces could be
implemented with an appropriate intensity mask, which determines
the relative amplitudes $\gamma_i$, and a global intensity
modulation, which gives the function $f(t)$. Expanding this
modulation in Fourier modes, we find $N(N-1)/2+N$ equations which
define a possible entangling procedure. We have solved numerically
these equations, both for the GHZ state and for the cluster state.
While in the first case we always found exact solutions with a
small number of modes (i.e. $50$ modes for $30$ ions), the
generation of the cluster state was always approximate with high
fidelity, $F \simeq 99.9\%$. The error in this case has its origin
in our particular choice of forces.

In Fig.~\ref{fig:entangling} we show the entangling procedure for a setup with
$10$ and $20$ ions, even though chains of up to $30$ ions have been
considered. We measure the fidelity of the process as the overlap between the
target state and the one achieved. If $\delta J$ is the difference between the
desired interaction and the achieved one, then $F = \frac{1}{2^N}
\sum_{\vec{s}} e^{-i\vec{s}^t \delta J T \vec{s}}$, where the sum is performed
over all possible spin configurations, $s_{k}=\pm 1$. The time scales for the
generation of the interaction are independent of the size of the system, and
for instance we can produce a GHZ state of $20$ ions in a time $T=1.1/\omega$,
to be compared with the time $T=3000\pi/\omega$ required when individually
addressing one of the vibrational modes \cite{molmer99}. The strength of the
forces, though, grows moderately with the number of ions, which can be
inconvenient.  However, thanks to the periodicity of the forcing, $f(t)$, if
we divide the intensity of the forces by a factor of 2, $f(t)\to f(t)/2$, the
same gate be produced in a longer time $4T$.  Furthermore, the forces that we
present in this paper are not optimal, and have been found with a
straightforward Gauss-Newton method. If high fidelity is not required, one may
find better solutions with fewer modes, but most important we expect
significant improvements by the application of better numerical algorithms to
search for the optimal entanglers.

Using the Ising interaction we can not only produce graph states, but also
squeezed states: states in which the variance of one spin component, $\Delta
S_x$, has been decreased at the expense of increasing the other variances. As
it was shown in \cite{kitagawa93}, a Hamiltonian of the form $H = J_z^2$
(single axis squeezing) or $H = J_x^2-J_y^2$ (two-axis squeezing) can be used
to produced squeezing. Both models can be simulated using our tools, either
directly, as in the single-axis squeezing, or stroboscopically, for the XY
interaction. Indeed, the stroboscopic simulation of the two-axis squeezing
resembles the scheme of $\pi/2$ pulses used in \cite{jaksch02} to effectively
switch off the interaction in two-mode Bose-Einstein condensates
\cite{sorensen01}.


\section{Optimal control of errors}
\label{sec:errors}

Up to now, we have assumed that the motion of the ions is not disturbed during
the time when the controlling forces are applied.  In this section we will
show how to take these effects into account for a realistic model of
dissipation. The main result is that the fidelity of the process can still be
computed and that the there are two sources of error: one due an imperfect
control of the ions, which introduces some temperature dependence on the
fidelity [Sec.~\ref{sec:temperature}], and another one due to the dissipation,
that can be treated as another constraint for the control problem
[Sec.~\ref{sec:fidelity-control}]. Finally, we will comment on possible
extensions outside the harmonic regime [Sec.~\ref{sec:duan}]

\subsection{The model and a exact solution}
\label{sec:master}

We study the dissipation with a master equation that arises from coupling the
phonon modes with a ``classical'' bosonic bath in thermal equilibrium
\begin{eqnarray}
  \label{master-eq}
  \frac{d}{dt}\rho &=& \frac{i}{\hbar}[H_I,\rho]
  +\sum_k \gamma N_k (2a_k^\dagger\rho a_k - a_ka_k^\dagger\rho-\rho
  a_ka_k^\dagger)\nonumber\\
  &+& \sum_k\gamma_k(N_k+1) (2a_k\rho a_k^\dagger - a_k^\dagger a_k\rho
  - \rho a_k^\dagger a_k).
\end{eqnarray}
Here $\gamma_k$ describes the coupling to an external bath and $N_k$ is the
mean number of bosons on that bath and it is related to its temperature. The
Hamiltonian in Eq.~(\ref{master-eq}) is written in the interaction picture
\begin{equation}
  H_I = \sum_{ki} \hbar(g_{ki}(t)a_k + g_{ki}^\star a_k^\dagger)\sigma^z_i.
\end{equation}
In order not to obscure the discussion, we will assume that each phonon mode
interacts with an independent bath. In that case, the forces in the rotating
frame of reference become, $\hbar g_{ki}(t) = \tfrac{1}{\sqrt{2}}F_i(t)
M_{ik}\alpha_k e^{-i\omega_k t}$. However, it is easy to generalize the
following analysis to a more realistic model in which each ion couples
independently to the environment, and the operators $a_k$ and $a_k^\dagger$ do
not represent the phonons, but the displacements of the ions.

To study the fidelity of a gate we only need to know the matrix elements
of the reduced density matrix for the internal degrees of freedom. This
matrix may be written as a collection of expectation values,
\begin{equation}
  \rho_{real} = \sum_{\mathbf{s},\mathbf{r}}
  \langle\Sigma_{\mathbf{s,r}}\rangle\Sigma_{\mathbf{s,r}},
\end{equation}
where $\Sigma_{\mathbf{s,r}} = |s_1\ldots s_N\rangle\langle r_1\ldots
r_N|$, $r_k,s_k=\pm 1$ form a complete basis for the space of
$2^N\times 2^N$ complex matrices.

The calculations that provide us with the evolution of
$\langle\Sigma_{\mathbf{s,r}}\rangle$ are detailed in Appendix
\ref{app:master}.  Here we will only summarize the main result, which is that
the reduced density matrix can be written as
\begin{equation}
  \langle \Sigma_{\mathbf{s},\mathbf{r}} \rangle(t) =
  e^{-\kappa_{\mathbf{s},\mathbf{r}}} e^{-i\sum_{jk}J_{jk} s_js_k}
  \langle \Sigma_{\mathbf{s},\mathbf{r}} \rangle(0) e^{i\sum_{jk}J_{jk}r_jr_k}.
\end{equation}
In other words, the spin density matrix has the form
\begin{equation}
  \label{sol-master-eq}
  \rho_{real}(t) = e^{-{\cal L}}\left[U \rho_{real}(0)U^\dagger\right],
\end{equation}
where $U=\exp(-i\sum J_{jk}\sigma^z_j\sigma^z_k)$ is the operation that we
want to perform, and ${\cal L}(\rho) = \sum_{\mathbf{s},\mathbf{r}}
\kappa_{\mathbf{s,r}}\Sigma_{\mathbf{s,r}}\rho\Sigma_{\mathbf{s,r}}$ is
responsible for the decay of coherences.

In comparison with the previous part of the paper, the unperturbed orbits in
phase space, that is the evolution without external forcing, are now not
circular orbits, but by spirally decaying ones. This fact translates in the
new conditions for uncoupling internal and motional degrees of freedom
(\ref{condition-N})
\begin{equation}
  \label{condition-1-dissip}
  \int_0^td\tau e^{-(i\omega_k+\gamma_k)\tau}F_j(\tau) = 0,\;\forall\,j,k
\end{equation}
which now depend on the exponential decay rate, $\gamma_k$, given by our
dissipation model. This model-dependence is also evident in the kernel that
produces our unitary operation, $J_{ij}$, which now reads
\begin{equation}
  \label{kernel-dissip}
  {\cal G}_{jl}(\tau) = \sum_{k} \frac{M_{jk}M_{lk}\alpha_k^2}{2m\omega_k\hbar}
  \sin(\omega_k|\tau|)e^{\gamma_k|\tau|}.
\end{equation}
Finally, we would like to remark that the conditions, the phase and the kernel
are only slightly modified when we consider a local coupling to the
environment.


\subsection{Quantum control of errors due to dissipation}
\label{sec:fidelity-control}

To understand better the implications of Eq.~(\ref{sol-master-eq}), let us
study the dynamics of a single ion. In this case the reduced density matrix
can be expressed uniquely in terms of the expectation values
$\langle\sigma^z\rangle$ and $\langle\sigma^+\rangle$. Furthermore, since the
magnetization is constant, we can compute the Uhlmann fidelity exactly as a
function of $\langle\sigma^+\rangle$.  Let us assume that initially the system
is in a pure state, and define $\langle \sigma^+\rangle_{real} =
e^{i\delta\phi - \kappa} \langle \sigma^+\rangle_{id}$, where the subindex
``id'' denotes the ideal value obtained in absence of errors. The Uhlmann
fidelity of the gate is \cite{nielsen} 
\begin{equation}
  {\cal F}(\rho_{real},\rho_{id}) =
  \sqrt{1 + |\langle\sigma^+\rangle_{id}|^2(2e^{-\kappa}\cos(\delta\phi)-1)}.
\end{equation}

Two types of errors contribute to the decay of the fidelity. The first type is
made of control errors. These errors contribute both to the spurious phases
($\delta\phi$) and to an effective decay due to not restoring the motional
state of the ions ($\kappa\neq0$ because $\beta(T)\neq 0$ and $U(T)\neq 1$).
These errors cause a smooth dependence on the temperature to appear, as shown
later in Sec.~\ref{sec:temperature}.

The second type of errors are due to dissipation during the forcing of the
ions. Their contribution to the exponential decay is
\begin{equation}
  \kappa_{\text{dissip}} = \gamma(N+\tfrac{1}{2}) \int_0^t d\tau|\beta(\tau)|^2.
\end{equation}
One would be led to think that if dissipation acts on a much larger scale than
the time required to perform our gate we can neglect it completely.  However,
a simple argument shows that this is not the case.  As we saw before in
Sec.~\ref{sec:continous}, the strength of the forces scales roughly as $F
\simeq T^{-5}$. This scaling allows us to give a worst case estimate of
$\beta(t) \simeq F t$ and conclude
\begin{equation}
  \kappa_{\text{dissip}} \sim \gamma T^{-2},\quad T\ll 1/\omega.
\end{equation}
What this means is that slower gates will involve smaller displacements of the
ions, which in turn translates into less dissipation. On the other hand, a too
long application of a force also gives more time for the dissipation to
operate and the optimal duration should be a compromise between both
processes. It is thus possible and recommended to \textit{optimize the forces
  $F(t)$} taking not only into account the properties of the force (i.e.
differentiablity and intensity), but also \textit{trying to minimize the decay
  $\kappa$ induced by the force}. From the numerical point of view, this new
control problem is only slightly more complicated than the ones we have solved
in Secs.~ \ref{sec:fast-2qb}-\ref{sec:control}, because
$\kappa_{\text{dissip}}$ is a nonlinear function of the forces.


\subsection{Errors due to an imperfect control: influence of temperature}
\label{sec:temperature}

Let us denote by $U_{id} = \exp(\sum_{ij}J_{ij}\sigma^z_i\sigma^z_j)$
the ideal operation that we want to perform, and by $U_{real}$ the
operation with errors. In this section, the only source of error that we
consider is an imperfect control, denoted by a perturbation of the
state-dependent force induced on the ions, $(F_i(t) + \delta
F_i(t))\sigma^z_i$. According to the previous analysis, the effect of
this perturbation will be a residual  state-dependent displacement of the
coherent wavepackets at the end of the process, $\beta_k(T)$,
\begin{equation}
  \beta_k = -i\int_0^T d\tau\,e^{i\omega_k \tau}
  \frac{\alpha_k M_{ik}\delta F_i(t)}{\hbar\sqrt{2}}\sigma^z_i
  =: \sum_{k,i}\beta_{ki}\sigma^z_i.
\end{equation}
plus a perturbation of the phase
\begin{equation}
  \delta \phi = \sum_{ij} \delta J_{ij}\sigma^z_i\sigma^z_j,
\end{equation}
which can be interpreted as a change in the effective interaction
between ions [Sec.~\ref{sec:control}]
\begin{eqnarray}
  \delta J_{ij} &=& \int_0^T\int_0^Td\tau_1d\tau_2\,{\cal
    G}_{ij}(\tau_2-\tau_1) \times\nonumber\\
  &&\times [F_i(\tau_1)\delta F_j(\tau_2)
  + \delta F_i(\tau_1) F_j(\tau_2)].\nonumber
\end{eqnarray}

We will assume as initial condition a pure state of the internal degrees
of freedom and a thermal state of the vibrational ones $\rho(0) =
|\psi\rangle\langle\psi|\otimes \rho_{vib}(T)$. The Uhlmann fidelity
at the end of the process is given by
\begin{eqnarray}
  {\cal F}(\rho_{id},\rho_{real}(t))
  &=& \mathrm{Tr}_{vib}(\langle\psi|U_{id}^\dagger\rho(t)U_{id}|\psi\rangle)
  \nonumber\\
  &=& \langle\psi|\mathrm{Tr}_{vib}[U_{id}^\dagger\rho(t)U_{id}]|\psi\rangle.
\end{eqnarray}
Expanding $|\psi\rangle= \sum c_{\mathbf{s}} |s_1\ldots s_N\rangle$, we obtain
\begin{equation}
  {\cal F} = \sum_{\mathbf{s},\mathbf{r}}
  c_{\mathbf{s}}^\star  c_{\mathbf{r}}
  \langle \Sigma_{\mathbf{s},\mathbf{r}}\rangle_{\rho(t)}.
\end{equation}
When we neglect dissipation, the previous expectation values can
be computed in terms of the final displacements, $\beta_k(t)$, and
the residual phases as follows:
\begin{eqnarray}
  \langle \Sigma_{\mathbf{s},\mathbf{r}}\rangle_{U_{id}^\dagger\rho(t)U_{id}}
  =  \mathrm{Tr}\left\{\prod_k D[\beta_k(t)(s_k-r_k)]\rho_{vib}(T)\right\}
  \times&&\\
  \times e^{i(\mathbf{s}^t\delta J\mathbf{s} - \mathbf{r}^t\delta J\mathbf{r})}
  c_{\mathbf{s}}c_{\mathbf{r}}^\star&&\nonumber
\end{eqnarray}
Here $D(z)=\exp(z a^\dagger - z^\star a)$ is the displacement operator,
$\rho_{vib}(T)$ is a thermal state
\begin{equation}
  \label{thermal-state}
  \rho_{vib}(T) = \bigotimes_{k=1}^N \frac{\hbar\omega_k}{\pi k_B T}
  \iint dzd\bar z\, e^{-|z|^2\hbar\omega_k/k_B T}|z\rangle_k\langle z|,
\end{equation}
and thus $\langle D(z)\rangle = \exp[-|z|^2(1/2+k_BT/\hbar\omega_k)] \equiv
\exp(C_k)$ so that the total fidelity becomes
\begin{eqnarray}
  \label{fid-thermal-1}
  {\cal F} = \sum_{\mathbf{s},\mathbf{r}}
  |c_{\mathbf{s}}|^2||c_{\mathbf{r}}|^2
  e^{i(\mathbf{s}^t\delta J\mathbf{s} - \mathbf{r}^t\delta J\mathbf{r})}
  e^{C_k (s_k-r_k)^2}.
\end{eqnarray}


\subsection{Errors due to larger displacements}
\label{sec:duan}

The previous studies can be generalized to arbitrary interactions and
trapping potentials. Let us assume a complicated Hamiltonian
\begin{equation}
  H = \sum_i \left[\frac{p_i^2}{2m} - f_i(t)\sigma^z_ix_i\right] +
  V(x_1,\ldots,x_N),
\end{equation}
describing the traps and the ion-ion interaction. The evolution equation
for the position of the ions are of the form
\begin{equation}
  \label{heisenberg}
  \dot x_i = \frac{p_k}{m},\;
  \dot p_i = -\frac{\partial V}{\partial x_i} + f_i(t)\sigma^z_i.
\end{equation}
Since the operators $\sigma^z_i$ are conserved quantities, the previous
equations can be thought of as a simple problem of Newtonian mechanics,
even though in practice, both $x_i$ and $p_i$ are operators. We can thus
represent a general solution as $\{x_i(t;\sigma),p_i(t;\sigma)\}$, where
$\sigma$ denotes the values of $\sigma^z_i$ operators.
The phase of the ions is then computed by analyzing the evolution of the
$\sigma^+_i$ operators. These operators must undergo a unitary
transformation $\sigma^+(t) = U(t)\sigma^+_i(0)U^\dagger(t)$ in which
the dependence on the $\{\sigma_i^z\}$ operators must be of the
form \footnote{We use the identity $(\sigma^z_j)^2=1$}
\begin{equation}
  U(t) = \exp[\sum_i \theta_i(t)\sigma^z_i +
  \sum_{ij}\phi_{ij}(t)\sigma^z_i\sigma^z_j] = e^{i\phi}.
\end{equation}
Using the commutation relation $[\sigma^+_i,\sigma^z_j] =
-2\delta_{ij}\sigma^z_i\sigma^+_i$, we obtain
\begin{equation}
  \sigma_i^+(t) = e^{2\theta_i\sigma^z_i+2\sum_{j\neq
  i}\phi_{ij}(t)\sigma^z_i\sigma^z_j}\sigma_i^+(0).
\end{equation}
Combining this with the Heisenberg equation for $\sigma_i^+$
\begin{equation}
  i\hbar\frac{d\sigma^+_i}{dt} = 2 f_i(t) x_i \sigma^z_i\sigma^+_i =
  2 \frac{\partial^2\phi}{\partial t\partial \sigma^z_i} \sigma^+_i,
\end{equation}
we find, up to global phases,
\begin{equation}
  \phi \sim \sum_i \int_0^T d\tau f_i(\tau)x_i(\tau)\sigma^z_i.
\end{equation}

From this analysis we see that we must impose two conditions on
the process. On one hand, the orbits of the ions must be periodic
so as to disentangle the internal and motional degrees of freedom
\begin{equation}
  \label{cond-heisenberg}
  x_i(T;\sigma)\simeq x_i(0),\; p_i(T;\sigma)\simeq p_i(0).
\end{equation}
On the other hand, the phase $\phi$ must be independent of the
initial conditions, $\{x_i(0), p_i(0)\}$. Satisfying both
conditions is impossible in general, but if we restrict ourselves
to small displacements and harmonic restoring forces, $\partial
V/\partial x_i \simeq \sum_j V_{ij}x_j$, it is possible to
integrate Eq.~(\ref{heisenberg}) and recover our expressions for
the phases (\ref{proto-ising}).

If, however, the qubit and higher terms in $V(x_1,\ldots,x_N)$ become
important, we will fail the restoring condition (\ref{cond-heisenberg}),
and induce some entanglement between the motion and the spin of the
ions. The errors due to these anharmonic terms are of the order
\begin{equation}
  \left|\frac{\partial^3 V}{\partial x_i\partial x_j\partial x_l}
    x_ix_jx_l\right|
  \sim \frac{3 \hbar \omega \alpha}{d} \left|\frac{x}{\alpha}\right|^3,
\end{equation}
where $\alpha \simeq \sqrt{\hbar/m\omega}$ is the length associated to
the harmonic oscillator, $d^3=e^2/2\pi\epsilon_0m\omega^2$ is the
equilibrium distance between two ions, and $x$ is a typical
displacement. Since a trivial analysis of these errors is not possible,
we can only produce a pessimistic, first order bound that restricts the
error induced by this perturbation on the wavefunction. First we will
give a worst case prediction for the maximum displacement of the ions as
$x_{max} < F_{max} T^2/2m$, where $F_{max}$ is the maximal force applied
on the ions. Next we will use the scaling $\phi \sim \omega^2 T^5
F_{max} / \hbar m$ to show that roughly $(x_{max}/\alpha) \sim
\phi/4\omega T$. With this, and first order perturbation theory we
compute the error and estimate it as
\begin{equation}
E_{anh} = \left(\frac{\alpha}{d}\right)^2
\frac{\phi^{3/2}}{4^{3/2} \omega T}.
\end{equation}
If we want to apply our phase operations to build a quantum
computer, we need $E < 10^{-4}$ and there is a limit on the speed
of the gate $T > 10^{-3}/\omega$, which nevertheless gives gating
rates of the order of 100 MHz.


\section{Conclusions}
\label{sec:conclusions}

We have developed a unified framework to study the coherent control of trapped
ions by means of state-dependent forces and robust geometric phases. Our
techniques can be used to perform fast two-qubit gates between pairs of ions.
For an adiabatic switching of the forces and for the case of pulsed lasers we
are able to reproduce the proposals of \cite{cirac00} and \cite{ripoll03}, and
with very little work we can design the optimal forces that produce a phase
gate in a given time with the lowest intensity.  Using the same tools and a
larger number of ions, we can simulate either continuously or stroboscopically
a number of spin $s=1/2$ Hamiltonians.  Furthermore, we are also able to
create highly entangled states and squeezing, and as prototypical examples we
have shown how to obtain a $GHZ$ state of $20$ ions in a very short time,
$T=1.1/\omega$. Finally we have studied the sources of error in the
application of our gate, which are an imperfect control, dissipation and
anharmonic terms in the interaction. The first type of errors could be ideally
corrected and introduce a smooth decay of the fidelity with the temperature.
The second type of errors induces also a decay of the fidelity, but the amount
of this error can be optimized using the tools of quantum control.  Both
dissipation and anharmonicity set up upper limits on the speed of the gate.
This limit is however very weak, since it allows theoretically a gating speed
of hundreds of MHz, and it could be overcome by a numerical study of the role
of anharmonic terms in the motion of the ions.

While concluding this paper we became aware of the work by P. Staanum, M.
Drewsen and K. M{\o}lmer \cite{staanum04} on performing quantum gates using
continuous laser beams. The ideas shown in Ref. \cite{staanum04} are
equivalent to the development of a two-qubit gate done in Sect.
\label{sec:geom-phase-2}, with the difference that we provide an optimal
solution for the problem.

J.J.G.-R. thanks J. Pachos, D. Porras and Shi-Liang Zhu for interesting
discussions during the development of this manuscript. Part of this work was
supported by the EU IST project RESQ, the EU project TOPQIP, the DFG
(Schwer\-punkt\-programm Quanten\-informations\-verarbeitung) and the
Kompetenz\-netz\-werk Quanten\-informations\-verarbeitung der Bayerischen
Staatsregierung.  Research at the University of Innsbruck is supported by the
Austrian Science Foundation, EU Networks and the Institute for Quantum
Information.

\appendix
\section{Solution of the master equation}
\label{app:master}

As we mentioned before, the density matrix is characterized by the expectation
values $\langle \Sigma_{\mathbf{s,r}}\rangle$. However, it is much more
difficult to work with these expectation values, than with
\begin{equation}
  \label{def-A}
  A := \Sigma_{\mathbf{s,r}} V(t) :=
  \Sigma_{\mathbf{s,r}} e^{\sum_k(\beta_ka_k^\dagger - \beta^\star_k a_k)}.
\end{equation}
By imposing that $V(0)=1$ and that $V(T)$ is at most a phase, we will be able
to relate the reduced density matrices $\rho_{real}(0)$ and $\rho_{real}(T)$.
It is easy to see that indeed the operator $V(t)$ is a displacement operator
and that Eq. (\ref{def-A}) is essentially the solution of the nondissipative
case, where $\beta_k$ measures the separation in phase space between
configurations with internal states $\mathbf{s}$ and $\mathbf{r}$.

The equation for the expectation value of an arbitrary operator, $A$, can be
written as follows
\begin{eqnarray}
  \frac{d}{dt}\langle{A}\rangle &=&
  \langle \partial_t A\rangle + \tfrac{i}\hbar\langle[H_I,A]\rangle
  - \gamma \langle (D_aA)a + a^\dagger(D_a)\rangle\nonumber\\
  &+& \gamma N\langle D_aD_{a^\dagger} A + D_{a^\dagger}D_a A\rangle.
  \label{moments}
\end{eqnarray}
Here, $D_aA := [A,a^\dagger]$ and $D_{a^\dagger}A:=[a,A]$ are two
superoperators which first of all commute, $[D_a,D_{a^\dagger}]=0$, and second
they are related to the formal derivatives with respect to the operators $a$
and $a^\dagger$. So, for instance, $D_a f(a,a^\dagger) = \partial_a
f(a,a^\dagger)$ for any analytical function $f$.

If we substitute Eq. (\ref{def-A}) into Eq. (\ref{moments}), and use
\begin{eqnarray}
  \partial_t A &=& \sum_k\frac{\partial A}{\partial\beta_k}\dot\beta_k
  + \sum_k\frac{\partial A}{\partial\beta_k^\star}\dot\beta_k^\star\nonumber\\
  &=& U\left[\dot\beta_k(a^\dagger_k - \tfrac{1}{2}\beta^\star_k)
    + \dot\beta_k^\star(-a_k-\tfrac{1}{2}\beta_k)\right],
\end{eqnarray}
we will obtain
\begin{eqnarray}
  \frac{d}{dt}\langle A\rangle &=&
  i\sum_k\left\langle A\left[(g_{k\mathbf{r}}-g_{k\mathbf{s}})a_k
      + g_{k\mathbf{r}}\beta_k + \mathrm{H.~c.}\right]\right\rangle\nonumber\\
  &+& \sum_k \left\langle A \left[\dot\beta_ka^\dagger_k - \dot\beta^\star_k
      a_k + \tfrac{1}{2}(\dot\beta_k\beta^\star_k - \beta_k\dot\beta_k^\star)
      \right]\right\rangle\nonumber\\
  &+& \sum_k\gamma_k N_k\left\langle A(-\beta_k^\star a_k+
    \beta_ka_k^\dagger)\right\rangle\nonumber\\
  &-&\sum_k\gamma_k\left(N_k+\tfrac{1}{2}\right)|\beta_k|^2\langle A\rangle,
\end{eqnarray}
whith the new parameters $g_{k\mathbf{r}} := \sum_i g_{ki}(t) r_i$.

Here is where we impose a particular evolution of the displacements on phase
space, $\dot\beta_k+\gamma_k\beta_k +i
(g_{k\mathbf{r}}^\star-g_{k\mathbf{s}}^\star)=0$. This equation has a trivial
solution
\begin{equation}
  \beta_k(t) = i \int_0^t d\tau\, e^{-\gamma_k(t-\tau)}
  [g_{k\mathbf{r}}^\star(\tau)-g_{k\mathbf{s}}^\star(\tau)].
\end{equation}
After substituting this value all terms containing Fock operators are
cancelled and we are left with
\begin{equation}
  \frac{d}{dt}\langle A\rangle = (-\dot\kappa + i\dot\phi) \langle A\rangle,
\end{equation}
where the decay $\kappa$ is
\begin{equation}
  \kappa(t) = \sum_k\gamma_k\left(N_k+\tfrac{1}{2}\right)
  \int_0^td\tau\,|\beta_k(\tau)|^2,
\end{equation}
and the total phase $\phi = \sum_{ij} J_{ij}(r_i+s_i)(r_j-s_j)$ is determined
by the matrix
\begin{equation}
  J_{ij} :=
  \mathrm{Im}\int_0^td\tau_1\int_0^{\tau_1}d\tau_2\,
  g_{ki}(\tau_1)g_{kj}(\tau_2)^\star e^{-\gamma_k(\tau_2-\tau_1)},
\end{equation}
Using the symmetry of this matrix, the formula for the phase can be rewritten
as $\phi = \sum_{ij} J_{ij}(r_ir_j-s_is_j)$, and the results mentioned in
Sect. \ref{sec:master} quickly follow.


\begin{thebibliography}{38}
\expandafter\ifx\csname natexlab\endcsname\relax\def\natexlab#1{#1}\fi
\expandafter\ifx\csname bibnamefont\endcsname\relax
  \def\bibnamefont#1{#1}\fi
\expandafter\ifx\csname bibfnamefont\endcsname\relax
  \def\bibfnamefont#1{#1}\fi
\expandafter\ifx\csname citenamefont\endcsname\relax
  \def\citenamefont#1{#1}\fi
\expandafter\ifx\csname url\endcsname\relax
  \def\url#1{\texttt{#1}}\fi
\expandafter\ifx\csname urlprefix\endcsname\relax\def\urlprefix{URL }\fi
\providecommand{\bibinfo}[2]{#2}
\providecommand{\eprint}[2][]{\url{#2}}

\bibitem[{\citenamefont{Cirac and Zoler}(2000)}]{cirac00}
\bibinfo{author}{\bibfnamefont{J.~I.} \bibnamefont{Cirac}} \bibnamefont{and}
  \bibinfo{author}{\bibfnamefont{P.}~\bibnamefont{Zoler}},
  \bibinfo{journal}{Nature} \textbf{\bibinfo{volume}{404}},
  \bibinfo{pages}{579} (\bibinfo{year}{2000}).

\bibitem[{\citenamefont{Garc{\'i}a-Ripoll
  et~al.}(2003)\citenamefont{Garc{\'i}a-Ripoll, Zoller, and Cirac}}]{ripoll03}
\bibinfo{author}{\bibfnamefont{J.~J.} \bibnamefont{Garc{\'i}a-Ripoll}},
  \bibinfo{author}{\bibfnamefont{P.}~\bibnamefont{Zoller}}, \bibnamefont{and}
  \bibinfo{author}{\bibfnamefont{J.~I.} \bibnamefont{Cirac}},
  \bibinfo{journal}{Phys. Rev. Lett.} \textbf{\bibinfo{volume}{91}},
  \bibinfo{pages}{157901} (\bibinfo{year}{2003}).

\bibitem[{\citenamefont{Levi}(2003)}]{levi03}
\bibinfo{author}{\bibfnamefont{B.~G.} \bibnamefont{Levi}},
  \bibinfo{journal}{Phys. Today} \textbf{\bibinfo{volume}{56}},
  \bibinfo{pages}{17} (\bibinfo{year}{2003}).

\bibitem[{\citenamefont{Monroe et~al.}(1995)\citenamefont{Monroe, Meekhof,
  King, Itano, and Wineland}}]{monroe95}
\bibinfo{author}{\bibfnamefont{C.}~\bibnamefont{Monroe}},
  \bibinfo{author}{\bibfnamefont{D.~M.} \bibnamefont{Meekhof}},
  \bibinfo{author}{\bibfnamefont{B.~E.} \bibnamefont{King}},
  \bibinfo{author}{\bibfnamefont{W.~M.} \bibnamefont{Itano}}, \bibnamefont{and}
  \bibinfo{author}{\bibfnamefont{D.~J.} \bibnamefont{Wineland}},
  \bibinfo{journal}{Phys. Rev. Lett.} \textbf{\bibinfo{volume}{75}},
  \bibinfo{pages}{4714} (\bibinfo{year}{1995}).

\bibitem[{\citenamefont{DeMarco et~al.}(2002)\citenamefont{DeMarco, Ben-KIsh,
  Leibfried, Meyer, Rowe, Jelenkovic, Itano, Briton, Langer, Rosenband
  et~al.}}]{demarco02}
\bibinfo{author}{\bibfnamefont{B.}~\bibnamefont{DeMarco}},
  \bibinfo{author}{\bibfnamefont{A.}~\bibnamefont{Ben-KIsh}},
  \bibinfo{author}{\bibfnamefont{D.}~\bibnamefont{Leibfried}},
  \bibinfo{author}{\bibfnamefont{V.}~\bibnamefont{Meyer}},
  \bibinfo{author}{\bibfnamefont{M.}~\bibnamefont{Rowe}},
  \bibinfo{author}{\bibfnamefont{B.~M.} \bibnamefont{Jelenkovic}},
  \bibinfo{author}{\bibfnamefont{W.~M.} \bibnamefont{Itano}},
  \bibinfo{author}{\bibfnamefont{J.}~\bibnamefont{Briton}},
  \bibinfo{author}{\bibfnamefont{C.}~\bibnamefont{Langer}},
  \bibinfo{author}{\bibfnamefont{T.}~\bibnamefont{Rosenband}},
  \bibnamefont{et~al.}, \bibinfo{journal}{Phys. Rev. Lett.}
  \textbf{\bibinfo{volume}{89}}, \bibinfo{pages}{267901}
  (\bibinfo{year}{2002}).

\bibitem[{\citenamefont{Leibfried et~al.}(2003)\citenamefont{Leibfried,
  DeMarco, Meyer, Lucas, Barret, Britton, Itano, Jelenkovic, Langer, Rosenband
  et~al.}}]{leibfried03}
\bibinfo{author}{\bibfnamefont{D.}~\bibnamefont{Leibfried}},
  \bibinfo{author}{\bibfnamefont{B.}~\bibnamefont{DeMarco}},
  \bibinfo{author}{\bibfnamefont{V.}~\bibnamefont{Meyer}},
  \bibinfo{author}{\bibfnamefont{D.}~\bibnamefont{Lucas}},
  \bibinfo{author}{\bibfnamefont{M.}~\bibnamefont{Barret}},
  \bibinfo{author}{\bibfnamefont{J.}~\bibnamefont{Britton}},
  \bibinfo{author}{\bibfnamefont{W.~M.} \bibnamefont{Itano}},
  \bibinfo{author}{\bibfnamefont{B.}~\bibnamefont{Jelenkovic}},
  \bibinfo{author}{\bibfnamefont{C.}~\bibnamefont{Langer}},
  \bibinfo{author}{\bibfnamefont{T.}~\bibnamefont{Rosenband}},
  \bibnamefont{et~al.}, \bibinfo{journal}{Nature}
  \textbf{\bibinfo{volume}{422}}, \bibinfo{pages}{412} (\bibinfo{year}{2003}).

\bibitem[{\citenamefont{Schmidt-Kaler et~al.}(2003)\citenamefont{Schmidt-Kaler,
  H{\"a}ffner, Riebe, Gulde, Lancaster, Deuschle, Becher, Roos, Eschner, and
  Blatt}}]{schmidtkaler03}
\bibinfo{author}{\bibfnamefont{F.}~\bibnamefont{Schmidt-Kaler}},
  \bibinfo{author}{\bibfnamefont{H.}~\bibnamefont{H{\"a}ffner}},
  \bibinfo{author}{\bibfnamefont{M.}~\bibnamefont{Riebe}},
  \bibinfo{author}{\bibfnamefont{S.}~\bibnamefont{Gulde}},
  \bibinfo{author}{\bibfnamefont{G.~P.~T.} \bibnamefont{Lancaster}},
  \bibinfo{author}{\bibfnamefont{T.}~\bibnamefont{Deuschle}},
  \bibinfo{author}{\bibfnamefont{C.}~\bibnamefont{Becher}},
  \bibinfo{author}{\bibfnamefont{C.~F.} \bibnamefont{Roos}},
  \bibinfo{author}{\bibfnamefont{J.}~\bibnamefont{Eschner}}, \bibnamefont{and}
  \bibinfo{author}{\bibfnamefont{R.}~\bibnamefont{Blatt}},
  \bibinfo{journal}{Nature} \textbf{\bibinfo{volume}{422}},
  \bibinfo{pages}{408} (\bibinfo{year}{2003}).

\bibitem[{\citenamefont{Kielpinski et~al.}(2002)\citenamefont{Kielpinski,
  Monroe, and Wineland}}]{kielpinski02}
\bibinfo{author}{\bibfnamefont{D.}~\bibnamefont{Kielpinski}},
  \bibinfo{author}{\bibfnamefont{C.}~\bibnamefont{Monroe}}, \bibnamefont{and}
  \bibinfo{author}{\bibfnamefont{D.~J.} \bibnamefont{Wineland}},
  \bibinfo{journal}{Nature} \textbf{\bibinfo{volume}{417}}
  (\bibinfo{year}{2002}).

\bibitem[{\citenamefont{Rowe et~al.}(2002)\citenamefont{Rowe, Ben-Kish,
  DeMarco, Leibfried, Meyer, Beall, Britton, Hughes, Itano, Jelenkovic
  et~al.}}]{rowe02}
\bibinfo{author}{\bibfnamefont{M.~A.} \bibnamefont{Rowe}},
  \bibinfo{author}{\bibfnamefont{A.}~\bibnamefont{Ben-Kish}},
  \bibinfo{author}{\bibfnamefont{B.}~\bibnamefont{DeMarco}},
  \bibinfo{author}{\bibfnamefont{D.}~\bibnamefont{Leibfried}},
  \bibinfo{author}{\bibfnamefont{V.}~\bibnamefont{Meyer}},
  \bibinfo{author}{\bibfnamefont{J.}~\bibnamefont{Beall}},
  \bibinfo{author}{\bibfnamefont{J.}~\bibnamefont{Britton}},
  \bibinfo{author}{\bibfnamefont{J.}~\bibnamefont{Hughes}},
  \bibinfo{author}{\bibfnamefont{W.~M.} \bibnamefont{Itano}},
  \bibinfo{author}{\bibfnamefont{B.}~\bibnamefont{Jelenkovic}},
  \bibnamefont{et~al.}, \bibinfo{journal}{Quantum Inf. and Comput.}
  \textbf{\bibinfo{volume}{2}}, \bibinfo{pages}{257} (\bibinfo{year}{2002}).

\bibitem[{\citenamefont{Wineland et~al.}(1992)\citenamefont{Wineland,
  Bollinger, Itano, Moore, and Heinzen}}]{wineland92}
\bibinfo{author}{\bibfnamefont{D.~J.} \bibnamefont{Wineland}},
  \bibinfo{author}{\bibfnamefont{J.~J.} \bibnamefont{Bollinger}},
  \bibinfo{author}{\bibfnamefont{W.~M.} \bibnamefont{Itano}},
  \bibinfo{author}{\bibfnamefont{F.~L.} \bibnamefont{Moore}}, \bibnamefont{and}
  \bibinfo{author}{\bibfnamefont{D.~J.} \bibnamefont{Heinzen}},
  \bibinfo{journal}{Phys. Rev. A} \textbf{\bibinfo{volume}{46}},
  \bibinfo{pages}{R6797} (\bibinfo{year}{1992}).

\bibitem[{\citenamefont{Wineland et~al.}(1994)\citenamefont{Wineland,
  Bollinger, Itano, and Heinzen}}]{wineland94}
\bibinfo{author}{\bibfnamefont{D.~J.} \bibnamefont{Wineland}},
  \bibinfo{author}{\bibfnamefont{J.~J.} \bibnamefont{Bollinger}},
  \bibinfo{author}{\bibfnamefont{W.~M.} \bibnamefont{Itano}}, \bibnamefont{and}
  \bibinfo{author}{\bibfnamefont{D.~J.} \bibnamefont{Heinzen}},
  \bibinfo{journal}{Phys. Rev. A} \textbf{\bibinfo{volume}{50}},
  \bibinfo{pages}{67} (\bibinfo{year}{1994}).

\bibitem[{\citenamefont{Steinbach and Gerry}(1998)}]{steinbach98}
\bibinfo{author}{\bibfnamefont{J.}~\bibnamefont{Steinbach}} \bibnamefont{and}
  \bibinfo{author}{\bibfnamefont{C.~C.} \bibnamefont{Gerry}},
  \bibinfo{journal}{Phys. Rev. Lett.} \textbf{\bibinfo{volume}{81}},
  \bibinfo{pages}{5528} (\bibinfo{year}{1998}).

\bibitem[{\citenamefont{M{\o}lmer and S{\o}rensen}(1999)}]{molmer99}
\bibinfo{author}{\bibfnamefont{K.}~\bibnamefont{M{\o}lmer}} \bibnamefont{and}
  \bibinfo{author}{\bibfnamefont{A.}~\bibnamefont{S{\o}rensen}},
  \bibinfo{journal}{Phys. Rev. Lett.} \textbf{\bibinfo{volume}{82}},
  \bibinfo{pages}{1835} (\bibinfo{year}{1999}).

\bibitem[{\citenamefont{S{\o}rensen and M{\o}lmer}(2000)}]{sorensen00}
\bibinfo{author}{\bibfnamefont{A.}~\bibnamefont{S{\o}rensen}} \bibnamefont{and}
  \bibinfo{author}{\bibfnamefont{K.}~\bibnamefont{M{\o}lmer}},
  \bibinfo{journal}{Phys. Rev. A} \textbf{\bibinfo{volume}{62}},
  \bibinfo{pages}{022311} (\bibinfo{year}{2000}).

\bibitem[{\citenamefont{Unanyan and Fleischhauer}(2003)}]{unanyan03}
\bibinfo{author}{\bibfnamefont{R.~G.} \bibnamefont{Unanyan}} \bibnamefont{and}
  \bibinfo{author}{\bibfnamefont{M.}~\bibnamefont{Fleischhauer}},
  \bibinfo{journal}{Phys. Rev. Lett.} \textbf{\bibinfo{volume}{90}},
  \bibinfo{pages}{133601} (\bibinfo{year}{2003}).

\bibitem[{\citenamefont{Turchette et~al.}(1998)\citenamefont{Turchette, Wood,
  King, Myatt, Leibfried, Itano, Monroe, and Wineland}}]{turchette98}
\bibinfo{author}{\bibfnamefont{Q.~A.} \bibnamefont{Turchette}},
  \bibinfo{author}{\bibfnamefont{C.~S.} \bibnamefont{Wood}},
  \bibinfo{author}{\bibfnamefont{B.~E.} \bibnamefont{King}},
  \bibinfo{author}{\bibfnamefont{C.~J.} \bibnamefont{Myatt}},
  \bibinfo{author}{\bibfnamefont{D.}~\bibnamefont{Leibfried}},
  \bibinfo{author}{\bibfnamefont{W.~M.} \bibnamefont{Itano}},
  \bibinfo{author}{\bibfnamefont{C.}~\bibnamefont{Monroe}}, \bibnamefont{and}
  \bibinfo{author}{\bibfnamefont{D.~J.} \bibnamefont{Wineland}},
  \bibinfo{journal}{Phys. Rev. Lett.} \textbf{\bibinfo{volume}{81}},
  \bibinfo{pages}{3631} (\bibinfo{year}{1998}).

\bibitem[{\citenamefont{Sackett et~al.}(2000)\citenamefont{Sackett, Kielpinski,
  King, Langer, Meyer, Myatt, Rowe, Turchette, Itano, Wineland
  et~al.}}]{sackett00}
\bibinfo{author}{\bibfnamefont{C.~A.} \bibnamefont{Sackett}},
  \bibinfo{author}{\bibfnamefont{D.}~\bibnamefont{Kielpinski}},
  \bibinfo{author}{\bibfnamefont{B.~E.} \bibnamefont{King}},
  \bibinfo{author}{\bibfnamefont{C.}~\bibnamefont{Langer}},
  \bibinfo{author}{\bibfnamefont{V.}~\bibnamefont{Meyer}},
  \bibinfo{author}{\bibfnamefont{C.~J.} \bibnamefont{Myatt}},
  \bibinfo{author}{\bibfnamefont{M.}~\bibnamefont{Rowe}},
  \bibinfo{author}{\bibfnamefont{Q.~A.} \bibnamefont{Turchette}},
  \bibinfo{author}{\bibfnamefont{W.~M.} \bibnamefont{Itano}},
  \bibinfo{author}{\bibfnamefont{D.~J.} \bibnamefont{Wineland}},
  \bibnamefont{et~al.}, \bibinfo{journal}{Nature}
  \textbf{\bibinfo{volume}{404}}, \bibinfo{pages}{256} (\bibinfo{year}{2000}).

\bibitem[{\citenamefont{Meyer et~al.}(2001)\citenamefont{Meyer, Rowe,
  Kielpinski, Sackett, Itano, Monroe, and Wineland}}]{meyer01}
\bibinfo{author}{\bibfnamefont{V.}~\bibnamefont{Meyer}},
  \bibinfo{author}{\bibfnamefont{M.~A.} \bibnamefont{Rowe}},
  \bibinfo{author}{\bibfnamefont{D.}~\bibnamefont{Kielpinski}},
  \bibinfo{author}{\bibfnamefont{C.~A.} \bibnamefont{Sackett}},
  \bibinfo{author}{\bibfnamefont{W.~M.} \bibnamefont{Itano}},
  \bibinfo{author}{\bibfnamefont{C.}~\bibnamefont{Monroe}}, \bibnamefont{and}
  \bibinfo{author}{\bibfnamefont{D.~J.} \bibnamefont{Wineland}},
  \bibinfo{journal}{Phys. Rev. Lett.} \textbf{\bibinfo{volume}{86}},
  \bibinfo{pages}{5870} (\bibinfo{year}{2001}).

\bibitem[{\citenamefont{Porras and Cirac}(2004)}]{porras04}
\bibinfo{author}{\bibfnamefont{D.}~\bibnamefont{Porras}} \bibnamefont{and}
  \bibinfo{author}{\bibfnamefont{J.~I.} \bibnamefont{Cirac}},
  \bibinfo{journal}{Phys. Rev. Lett.} \textbf{\bibinfo{volume}{92}},
  \bibinfo{pages}{207901} (\bibinfo{year}{2004}).

\bibitem[{\citenamefont{Barjaktarevic et~al.}()\citenamefont{Barjaktarevic,
  Milburn, and McKenzie}}]{barjaktarevic}
\bibinfo{author}{\bibfnamefont{J.}~\bibnamefont{Barjaktarevic}},
  \bibinfo{author}{\bibfnamefont{G.~J.} \bibnamefont{Milburn}},
  \bibnamefont{and} \bibinfo{author}{\bibfnamefont{R.~H.}
  \bibnamefont{McKenzie}}, \bibinfo{note}{arXive:quant-ph/0401137}.

\bibitem[{\citenamefont{Cirac and Zoller}(1995)}]{cirac95}
\bibinfo{author}{\bibfnamefont{J.~I.} \bibnamefont{Cirac}} \bibnamefont{and}
  \bibinfo{author}{\bibfnamefont{P.}~\bibnamefont{Zoller}},
  \bibinfo{journal}{Phys. Rev. Lett.} \textbf{\bibinfo{volume}{74}},
  \bibinfo{pages}{4091} (\bibinfo{year}{1995}).

\bibitem[{\citenamefont{Poyatos et~al.}(1998)\citenamefont{Poyatos, Cirac, and
  Zoller}}]{poyatos98}
\bibinfo{author}{\bibfnamefont{J.~F.} \bibnamefont{Poyatos}},
  \bibinfo{author}{\bibfnamefont{J.~I.} \bibnamefont{Cirac}}, \bibnamefont{and}
  \bibinfo{author}{\bibfnamefont{P.}~\bibnamefont{Zoller}},
  \bibinfo{journal}{Phys. Rev. Lett.} \textbf{\bibinfo{volume}{81}},
  \bibinfo{pages}{1322} (\bibinfo{year}{1998}).

\bibitem[{\citenamefont{S{\o}rensen and M{\o}lmer}(1999)}]{sorensen99}
\bibinfo{author}{\bibfnamefont{A.}~\bibnamefont{S{\o}rensen}} \bibnamefont{and}
  \bibinfo{author}{\bibfnamefont{K.}~\bibnamefont{M{\o}lmer}},
  \bibinfo{journal}{Phys. Rev. Lett.} \textbf{\bibinfo{volume}{82}},
  \bibinfo{pages}{1971} (\bibinfo{year}{1999}).

\bibitem[{\citenamefont{Jonathan et~al.}(2000)\citenamefont{Jonathan, Plenio,
  and Knight}}]{jonathan00}
\bibinfo{author}{\bibfnamefont{D.}~\bibnamefont{Jonathan}},
  \bibinfo{author}{\bibfnamefont{M.~B.} \bibnamefont{Plenio}},
  \bibnamefont{and} \bibinfo{author}{\bibfnamefont{P.~L.}
  \bibnamefont{Knight}}, \bibinfo{journal}{Phys. Rev. A}
  \textbf{\bibinfo{volume}{62}}, \bibinfo{pages}{042307}
  (\bibinfo{year}{2000}).

\bibitem[{\citenamefont{James}(2000)}]{james00}
\bibinfo{author}{\bibfnamefont{D.~F.~V.} \bibnamefont{James}},
  \bibinfo{journal}{Fortschr. Phys.} \textbf{\bibinfo{volume}{48}},
  \bibinfo{pages}{9} (\bibinfo{year}{2000}).

\bibitem[{\citenamefont{Milburn et~al.}(2000)\citenamefont{Milburn, Schneider,
  and James}}]{milburn00}
\bibinfo{author}{\bibfnamefont{G.~J.} \bibnamefont{Milburn}},
  \bibinfo{author}{\bibfnamefont{S.}~\bibnamefont{Schneider}},
  \bibnamefont{and} \bibinfo{author}{\bibfnamefont{D.~F.~V.}
  \bibnamefont{James}}, \bibinfo{journal}{Fortschr. Phys.}
  \textbf{\bibinfo{volume}{48}}, \bibinfo{pages}{9} (\bibinfo{year}{2000}).

\bibitem[{\citenamefont{Duan}()}]{duan04}
\bibinfo{author}{\bibfnamefont{L.-M.} \bibnamefont{Duan}},
  \bibinfo{note}{arXiv:quant-ph/041185}.

\bibitem[{\citenamefont{Staanum et~al.}(2004)\citenamefont{Staanum, Drewsen,
  and M{\o}lmer}}]{staanum04}
\bibinfo{author}{\bibfnamefont{P.}~\bibnamefont{Staanum}},
  \bibinfo{author}{\bibfnamefont{M.}~\bibnamefont{Drewsen}}, \bibnamefont{and}
  \bibinfo{author}{\bibfnamefont{K.}~\bibnamefont{M{\o}lmer}},
  \bibinfo{journal}{Phys. Rev. A} \textbf{\bibinfo{volume}{70}},
  \bibinfo{pages}{052327} (\bibinfo{year}{2004}).

\bibitem[{\citenamefont{Aharonov and Anandan}(1987)}]{aharonov87}
\bibinfo{author}{\bibfnamefont{Y.}~\bibnamefont{Aharonov}} \bibnamefont{and}
  \bibinfo{author}{\bibfnamefont{J.}~\bibnamefont{Anandan}},
  \bibinfo{journal}{Phys. Rev. Lett.} \textbf{\bibinfo{volume}{58}},
  \bibinfo{pages}{1593} (\bibinfo{year}{1987}).

\bibitem[{\citenamefont{Calarco et~al.}(2001)\citenamefont{Calarco, Cirac, and
  Zoller}}]{calarco01}
\bibinfo{author}{\bibfnamefont{T.}~\bibnamefont{Calarco}},
  \bibinfo{author}{\bibfnamefont{J.~I.} \bibnamefont{Cirac}}, \bibnamefont{and}
  \bibinfo{author}{\bibfnamefont{P.}~\bibnamefont{Zoller}},
  \bibinfo{journal}{Phys. Rev. A} \textbf{\bibinfo{volume}{63}},
  \bibinfo{pages}{062304} (\bibinfo{year}{2001}).

\bibitem[{\citenamefont{Sasura and Steane}(2003)}]{sasura03}
\bibinfo{author}{\bibfnamefont{M.}~\bibnamefont{Sasura}} \bibnamefont{and}
  \bibinfo{author}{\bibfnamefont{A.~M.} \bibnamefont{Steane}},
  \bibinfo{journal}{Phys. Rev. A} \textbf{\bibinfo{volume}{67}},
  \bibinfo{pages}{062318} (\bibinfo{year}{2003}).

\bibitem[{\citenamefont{Hein et~al.}(2004)\citenamefont{Hein, Eisert, and
  Briegel}}]{hein04}
\bibinfo{author}{\bibfnamefont{M.}~\bibnamefont{Hein}},
  \bibinfo{author}{\bibfnamefont{J.}~\bibnamefont{Eisert}}, \bibnamefont{and}
  \bibinfo{author}{\bibfnamefont{H.~J.} \bibnamefont{Briegel}},
  \bibinfo{journal}{Phys. Rev. Lett.} \textbf{\bibinfo{volume}{69}},
  \bibinfo{pages}{062311} (\bibinfo{year}{2004}).

\bibitem[{\citenamefont{Briegel and Raussendorf}(2001)}]{briegel01}
\bibinfo{author}{\bibfnamefont{H.~J.} \bibnamefont{Briegel}} \bibnamefont{and}
  \bibinfo{author}{\bibfnamefont{R.}~\bibnamefont{Raussendorf}},
  \bibinfo{journal}{Phys. Rev. Lett.} \textbf{\bibinfo{volume}{86}},
  \bibinfo{pages}{910} (\bibinfo{year}{2001}).

\bibitem[{\citenamefont{Raussendorf and Briegel}(2001)}]{raussendorf01}
\bibinfo{author}{\bibfnamefont{R.}~\bibnamefont{Raussendorf}} \bibnamefont{and}
  \bibinfo{author}{\bibfnamefont{H.~J.} \bibnamefont{Briegel}},
  \bibinfo{journal}{Phys. Rev. Lett.} \textbf{\bibinfo{volume}{86}},
  \bibinfo{pages}{5188} (\bibinfo{year}{2001}).

\bibitem[{\citenamefont{Kitagawa and Ueda}(1993)}]{kitagawa93}
\bibinfo{author}{\bibfnamefont{M.}~\bibnamefont{Kitagawa}} \bibnamefont{and}
  \bibinfo{author}{\bibfnamefont{M.}~\bibnamefont{Ueda}},
  \bibinfo{journal}{Phys. Rev. A} \textbf{\bibinfo{volume}{47}},
  \bibinfo{pages}{5138} (\bibinfo{year}{1993}).

\bibitem[{\citenamefont{Jaksch et~al.}(2002)\citenamefont{Jaksch, Cirac, and
  Zoller}}]{jaksch02}
\bibinfo{author}{\bibfnamefont{D.}~\bibnamefont{Jaksch}},
  \bibinfo{author}{\bibfnamefont{J.~I.} \bibnamefont{Cirac}}, \bibnamefont{and}
  \bibinfo{author}{\bibfnamefont{P.}~\bibnamefont{Zoller}},
  \bibinfo{journal}{Phys. Rev. A} \textbf{\bibinfo{volume}{65}},
  \bibinfo{pages}{033625} (\bibinfo{year}{2002}).

\bibitem[{\citenamefont{Sorensen et~al.}(2001)\citenamefont{Sorensen, Duan,
  Cirac, and Zoller}}]{sorensen01}
\bibinfo{author}{\bibfnamefont{A.}~\bibnamefont{Sorensen}},
  \bibinfo{author}{\bibfnamefont{L.-M.} \bibnamefont{Duan}},
  \bibinfo{author}{\bibfnamefont{J.~I.} \bibnamefont{Cirac}}, \bibnamefont{and}
  \bibinfo{author}{\bibfnamefont{P.}~\bibnamefont{Zoller}},
  \bibinfo{journal}{Nature} \textbf{\bibinfo{volume}{403}}, \bibinfo{pages}{63}
  (\bibinfo{year}{2001}).

\bibitem[{\citenamefont{Nielsen and Chuang}(2000)}]{nielsen}
\bibinfo{author}{\bibfnamefont{M.~A.} \bibnamefont{Nielsen}} \bibnamefont{and}
  \bibinfo{author}{\bibfnamefont{I.~L.} \bibnamefont{Chuang}},
  \emph{\bibinfo{title}{Quantum Computation and Quantum Information}}
  (\bibinfo{publisher}{Cambridge Univ. Press}, \bibinfo{address}{Cambridge},
  \bibinfo{year}{2000}).

\end{thebibliography}
\end{document}